\def\omitcomments{1}
\newif\ifcamera
\documentclass[letterpaper,twocolumn,10pt]{article}
\usepackage{usenix2019_v3}
\ifcamera
\usepackage[available,functional]{usenixbadges}
\fi

\ifcamera
\fi

\usepackage{latexsym,amsmath,amsfonts,amssymb,stmaryrd,mathtools}
\usepackage{amsthm}
\usepackage{algorithm, algorithmicx}
\usepackage[noend]{algpseudocode}
\usepackage{graphicx}
\usepackage{hyperref}
\usepackage{wrapfig}
\usepackage{listings, fancyvrb}
\usepackage{multirow}
\usepackage{comment}
\usepackage{xspace}
\usepackage{pifont}
\usepackage{parcolumns}
\usepackage{graphicx}
\usepackage{enumerate}
\usepackage{enumitem}
\usepackage{wrapfig}
\usepackage{tabularx}
\usepackage{titlecaps}
\usepackage{subcaption}
\usepackage{tikz}
\usepackage{pgfplots}
\usepackage{booktabs}
\microtypecontext{spacing=nonfrench}

\newtheorem{theorem}{Theorem}[section]
\newtheorem{lemma}[theorem]{Lemma}
\newtheorem{corollary}[theorem]{Corollary}
\newtheorem{invariant}[theorem]{Invariant}
\newtheorem{definition}{Definition}

\newcommand\rmv[1]{}
\def\eg{e.g.}

\newcommand{\node}{replica}
\newcommand{\rplane}{replication plane}
\newcommand{\leplane}{background plane}
\newcommand{\sysname}{Mu\xspace}
\newcommand{\redis}{Redis\xspace}
\newcommand{\memcached}{Memcached\xspace}

\if\omitcomments 1
 \newcommand{\Naama}[1]{}
 \newcommand{\igor}[1]{}
 \newcommand{\mka}[1]{}
 \newcommand{\vjm}[1]{}
 
\else
 \newcommand{\Naama}[1]{\noindent
 	{\color{blue}{\textbf{Naama: }#1}}}
 \newcommand{\igor}[1]{\noindent
    {\color{brown}{\textbf{Igor: }#1}}}
 \newcommand{\mka}[1]{\noindent
    {\color{orange}{\textbf{MKA: }#1}}}
 \newcommand{\vjm}[1]{\noindent
    {\color{red}{\textbf{VJM: }#1}}}
 
\fi
\newcommand{\CR}[1]{{#1}}

\newcommand{\CRExtra}[1]{{#1}}

\makeatletter
\lst@Key{countblanklines}{true}[t]%
{\lstKV@SetIf{#1}\lst@ifcountblanklines}

\lst@AddToHook{OnEmptyLine}{%
	\lst@ifnumberblanklines\else%
	\lst@ifcountblanklines\else%
	\advance\c@lstnumber-\@ne\relax%
	\fi%
	\fi}
\makeatother

\graphicspath{{../figures/}}

\usetikzlibrary{
    pgfplots.colorbrewer,
} 
\pgfplotsset{compat=newest}
\pgfplotsset{
    tick label style={font=\small},
    label style={font=\small},
    legend style={font=\footnotesize},
    every axis/.append style={line width=1pt},
    cycle list name=exotic,
    cycle list/.define={my marks}{
        every mark/.append style={solid,fill=\pgfkeysvalueof{/pgfplots/mark list fill}},mark=*\\
        every mark/.append style={solid,fill=\pgfkeysvalueof{/pgfplots/mark list fill}},mark=square*\\
        every mark/.append style={solid,fill=\pgfkeysvalueof{/pgfplots/mark list fill}},mark=triangle*\\
        every mark/.append style={solid,fill=\pgfkeysvalueof{/pgfplots/mark list fill}},mark=diamond*\\
    },
    mark list fill={.!75!white},
    cycle multiindex* list={
        exotic
            \nextlist
        my marks
            \nextlist
        linestyles*
            \nextlist
        very thick
            \nextlist
    },
}
\usepgfplotslibrary{groupplots}

\begin{document}


\title{\resizebox{\textwidth}{!}{\mbox{Microsecond Consensus for Microsecond Applications}}}


\author{
{\rm Marcos K. Aguilera}\\
VMware Research
\and
{\rm Naama Ben-David}\\
VMware Research
\and
{\rm Rachid Guerraoui}\\
EPFL
\and
{\rm Virendra J. Marathe}\\
Oracle Labs
\and
{\rm Athanasios Xygkis}\\
EPFL
\and
{\rm Igor Zablotchi}\\
EPFL
} 

\maketitle

\begin{abstract}
We consider the problem of making apps fault-tolerant through replication, when apps operate at the microsecond scale, as in finance, embedded computing, and microservices apps. These apps need a replication scheme that also operates at the microsecond scale, otherwise replication becomes a burden. We propose \sysname, a system that takes less than 1.3 microseconds to replicate a (small) request in memory, and less than a millisecond
to fail-over the system---this cuts the replication and fail-over latencies of the
prior systems by at least 61\% and 90\%.
 \sysname implements bona fide state machine replication/consensus (SMR) with strong consistency for a generic app, but it really shines on microsecond apps, where even the smallest overhead is significant. To provide this performance, \sysname introduces a new SMR protocol that carefully leverages RDMA. Roughly, in \sysname a leader replicates a request by simply writing it directly to the log of other replicas using RDMA, without any additional communication. Doing so, however, introduces the challenge of handling concurrent leaders, changing leaders, garbage collecting the logs, and more---challenges that we address in this paper through a judicious combination of RDMA permissions and distributed algorithmic design.
We implemented \sysname and used it to replicate several
  systems: a financial exchange app called Liquibook,
  \redis{}, \memcached{},
  and HERD~\cite{kalia2014using}.
Our evaluation shows that \sysname incurs a small replication latency, in some
  cases being the only viable replication system that incurs an acceptable overhead.
\end{abstract}

\section{Introduction}




Enabled 
by modern technologies such as RDMA, Microsecond-scale computing is emerging as a must~\cite{mukiller}.
A microsecond app might be expected to process a request in 10 microseconds.
Areas where software systems care about microsecond performance include finance 
(\eg, trading systems), embedded computing (\eg, control systems), and microservices (\eg, key-value stores).
Some of these areas are critical and it is desirable to replicate their microsecond apps 
across many hosts to provide high availability, due to economic, safety, 
or robustness reasons.
Typically, a system may have hundreds of microservice apps~\cite{mubench}, some of which are stateful 
and can disrupt a global execution if they fail (\eg, key-value stores)---these apps should be replicated for 
the sake of the whole system.
%
%
%

The golden standard to replicate an app is State Machine Replication (SMR)~\cite{schneider1990implementing}, whereby
replicas execute requests in the same total order determined by a consensus protocol. 
Unfortunately, traditional SMR systems add hundreds of microseconds of overhead even on a fast network~\cite{hunt2010zookeeper}.
Recent work explores modern hardware in order to improve the performance of replication~\cite{wang2017apus,poke2015dare,hermes,erpc,derecho,hovercraft}. 
The fastest of these (\eg, Hermes~\cite{hermes}, DARE~\cite{poke2015dare}, and HovercRaft~\cite{hovercraft}) induce however 
an overhead of several microseconds, which is 
clearly high for apps that themselves take few microseconds.
Furthermore, when a failure occurs, prior systems incur a prohibitively large fail-over time in the tens 
of \emph{milli}seconds (not microseconds). For instance, HovercRaft takes 10 milliseconds,
DARE 30 milliseconds, and Hermes at least 150 milliseconds. 
The rationale for such large latencies are timeouts that account for the natural fluctuations in the latency of modern networks.
Improving replication and fail-over latencies requires fundamentally new techniques.



We propose \sysname, a new SMR system that adds less than 1.3 microseconds to replicate a (small) app request, 
with the 99th-percentile at 1.6 microseconds.
Although \sysname is a general-purpose SMR scheme for a generic app, \sysname really shines with microsecond apps, 
where even the smallest replication overhead is significant.
Compared to the fastest prior system, \sysname is able to cut 61\% of its latency.
This is the smallest latency possible with current RDMA hardware, as it corresponds to one round of \emph{one-sided} communication.

To achieve this performance, \sysname introduces a new SMR protocol that fundamentally changes how RDMA can be leveraged for replication. 
Our protocol reaches consensus and replicates a request with just one round of
parallel RDMA write operations on a majority of replicas.
This is in contrast to prior approaches, which take multiple rounds~\cite{derecho,poke2015dare,wang2017apus} or resort to two-sided communication~\cite{hunt2010zookeeper,erpc,kotla2007zyzzyva,mazieres2007paxos}. Roughly, in \sysname the leader replicates a request by simply using RDMA to write it to the log of each replica, without additional rounds of communication. 
Doing this correctly is challenging because concurrent leaders may try to write to the logs simultaneously. 
In fact, the hardest part of most replication protocols is the mechanism to protect against races of concurrent 
leaders (\eg, Paxos proposal numbers~\cite{paxos}). 
Traditional replication implements this mechanism using send-receive communication (two-sided operations) or multiple rounds of communication.
Instead, \sysname uses RDMA write permissions to guarantee that a replica's log can be written by only one leader. 
Critical to correctness are the mechanisms to change leaders and garbage collect logs, as we describe in the paper.

\sysname also improves fail-over time to just 873 microseconds, with the 99-th percentile at 945 microseconds,
which cuts fail-over time of prior systems by an order of magnitude.
%
%
%
The fact that \sysname significantly improves both replication overhead and fail-over latency is perhaps surprising: folklore suggests a trade-off between the latencies of replication in the fast path, and fail-over in the slow path.

The fail-over time of \sysname has two parts: failure detection and leader change.
%
%
%
%
For failure detection, traditional SMR systems typically use a timeout on heartbeat messages from the leader. Due to large variances in network latencies, timeout values are in the 10--100ms even with the fastest networks. This is clearly high for microsecond apps. \sysname uses a conceptually different method based on a pull-score mechanism over RDMA. The leader increments a heartbeat counter in its local memory, while other replicas use RDMA to periodically read the counter and calculate a badness score. The score is the number of successive reads that returned the same value. Replicas declare a failure if the score is above a threshold, corresponding to a timeout. Different from the traditional heartbeats, this method can use an aggressively small timeout without false positives because network delays slow down the reads rather than the heartbeat. In this way, \sysname detects failures usually within ${\sim}$600 microseconds.
This is bottlenecked
by variances in process scheduling, as we discuss later.

For leader change, the latency comes from the cost of changing RDMA write permissions, which with current NICs are hundreds of microseconds. This is higher than we expected: it is far slower than RDMA reads and writes, which go over the network. We attribute this delay to a lack of hardware optimization. RDMA has many methods to change permissions: (1) re-register memory regions, (2) change queue-pair access flags, or (3) close and reopen queue pairs. We carefully evaluate the speed of each method and propose a scheme that combines two of them using a fast-slow path to minimize latency. Despite our efforts, the best way to cut this latency further is to improve the NIC hardware.

We prove that 
\sysname provides strong consistency in the form of linearizability~\cite{HW90}, despite crashes and 
asynchrony, and it ensures liveness under the
same assumptions as Paxos~\cite{paxos}.

We implemented \sysname
and used it to replicate several apps:
  a financial exchange app called Liquibook~\cite{liquibook},
  \redis{}, \memcached{},
  and an RDMA-based key-value stored called HERD~\cite{kalia2014using}.
  
We evaluate \sysname extensively, by studying its
  replication latency stand-alone or integrated into each of the
  above apps.
We find that, for some of these apps (Liquibook, HERD), 
  \sysname is the only viable replication system that incurs a reasonable overhead.
This is because \sysname's latency is significantly lower by a factor of at least
   2.7${\times}$ compared to other replication systems.
We also report on our study of \sysname's fail-over latency, with a breakdown
  of its components, suggesting ways to improve
  the infrastructure to further reduce the latency.
  

\sysname has some limitations.
First, \sysname relies on RDMA and so it is suitable only for networks with RDMA, 
  such as local area networks, but not across the wide area.
Second, \sysname is an in-memory system
  that does not persist data in
  stable storage---doing so would add additional latency
  dependent on the device speed.
 \footnote{For fairness, all SMR systems that
   we compare against also operate in-memory.}
However, we observe that the industry is working on extensions of RDMA for persistent memory, whereby RDMA writes can be flushed at a remote persistent memory with minimum latency~\cite{rdmapmem}---once available, this extension
  will provide persistence for \sysname.


To summarize, we make the following contributions:

\begin{itemize}
\item We propose \sysname, a new SMR system with low replication and fail-over latencies.

\item To achieve its performance, \sysname leverages RDMA permissions and a scoring mechanism over heartbeat counters.


\item We give the complete correctness proof of \sysname.
\item We implement \sysname, and evaluate both its raw performance and its performance in microsecond apps. Results show that \sysname significantly reduces replication latencies to an acceptable level for microsecond apps. 
\item \sysname{'s} code is available at:\\ \url{https://github.com/LPD-EPFL/mu}.
\end{itemize}

One might argue that \sysname is ahead of its time, as
  most apps today are not yet microsecond apps.
However, this situation is changing.
We already have important microsecond apps in areas
  such as trading, and more will come
  as existing timing requirements
  become stricter and new systems emerge as
  the composition of a large number of 
  microservices (\S\ref{sec:mu}).

\section{Background} \label{sec:background}

\subsection{Microsecond Apps and Computing}\label{sec:mu}

Apps that are consumed by humans typically work at the millisecond scale: to the 
  human brain, the lowest reported perceptible latency is 13 milliseconds~\cite{rsvp}.
Yet, we see the emergence of apps that
  are consumed not by humans but by 
  other computing systems.
An increasing number of such systems must
  operate at the microsecond scale, for competitive,
  physical, or composition reasons.
Schneider~\cite{mumarket} speaks of a microsecond market where traders spend massive resources to gain a microsecond advantage in their high-frequency trading.
Industrial robots must orchestrate their motors with microsecond granularity for precise movements~\cite{mumotor}.
Modern distributed systems are composed of
  hundreds~\cite{mubench}
  of
  stateless and
  stateful microservices, such as
  key-value stores, web servers, load balancers, and
  ad services---each operating
  as an independent app whose
  latency requirements are gradually decreasing
  to the microsecond level~\cite{muback},
  as the number of composed services is
  increasing.
With this trend, we already see the emergence
  of key-value stores with microsecond latency (\eg, \cite{storm,erpc}).

To operate at the microsecond scale, the computing
  ecosystem must be improved at many layers.
This is happening gradually by various recent efforts.
Barroso et al~\cite{mukiller} argue for
  better support of microsecond-scale
  events.
The latest Precision Time Protocol improves
  clock synchronization to
  achieve submicrosecond accuracy~\cite{ptplatest}.
And other recent work improves 
  CPU scheduling~\cite{muback,shenango,zygos},
  thread management~\cite{arachne},
  power management~\cite{ix},
  RPC handling~\cite{erpc,rpcvalet},
  and the network stack~\cite{shenango}---all
  at the microsecond scale.
\sysname fits in this context, by providing
   microsecond SMR.

\subsection{State Machine Replication}
State Machine Replication (SMR) replicates a service (\eg, a key-value storage system)
  across multiple physical servers called \emph{replicas}, such that the system remains available and consistent even if some servers fail.
SMR provides strong consistency in the form of linearizability~\cite{HW90}.
A common way to implement SMR, which we adopt in this paper, is as follows: each replica
  has a copy of the service software and a log. 
The log stores client requests.
We consider non-durable SMR
  systems~\cite{ramcloud,li2016just,curp,jin2018netchain,istvan2016consensus,ipipe},
  which keep state in memory only, without logging updates to stable storage. 
Such systems make an item of data reliable by keeping copies of it in the memory of several nodes. Thus, the data remains recoverable as long as there are fewer simultaneous node failures than data copies~\cite{poke2015dare}.

A consensus protocol ensures that all replicas agree on what request is stored in each 
  slot of the log.
Replicas then apply the requests in the log (i.e., execute the 
corresponding operations), in log order.
Assuming that the service is deterministic, this ensures all replicas remain
  in sync.
We adopt a leader-based approach, in which a dynamically elected replica 
  called the \emph{leader} communicates with the clients and sends back responses
  after requests reach a majority of replicas.
We assume a \textit{crash-failure} model: servers may fail by crashing, after which they
  stop executing. 

A consensus protocol must ensure \textit{safety} and \textit{liveness} properties. 
Safety here means 
(1) \textit{agreement} (different replicas do not obtain different values for
   a given log slot)
   and
(2) \textit{validity} (replicas do not obtain spurious values).
Liveness means \textit{termination}---every live replica eventually obtains a value.
\CR{We guarantee agreement and validity in an asynchronous system, while termination 
  requires eventual synchrony and a majority of non-crashed replicas, as in typical consensus protocols.
  In theory, it is possible to design systems that terminate under weaker synchrony~\cite{chandra1996weakest}, 
  but this is not our goal.}


\subsection{RDMA}

Remote Direct Memory Access (RDMA) allows a host to access the memory of another host
  without involving the processor at the other host.
RDMA enables low-latency communication by bypassing the OS kernel and by implementing several layers of the network stack in hardware.

RDMA supports many operations: Send/Receive, Write/Read, and Atomics 
  (compare-and-swap, fetch-and-increment). 
  Because of their lower latency, we use only RDMA Writes and Reads.
RDMA has several transports; we use Reliable Connection (RC) to provide in-order reliable delivery.
%

RDMA connection endpoints are called Queue Pairs (QPs). 
Each QP is associated to a Completion Queue (CQ). 
Operations are posted to QPs as Work Requests (WRs). 
The RDMA hardware consumes the WR, performs the operation, and posts a Work Completion (WC) to the CQ. 
Applications make local memory available for remote access by registering local virtual memory regions (MRs) with the RDMA driver. Both QPs and MRs can have different access modes
  (e.g., read-only or read-write).
The access mode is specified when initializing the QP or registering the MR, but can be changed later.
MRs can overlap: the same memory can be registered multiple times, yielding multiple MRs,
  each with its own access mode. 
In this way, different remote machines can have different access rights to the same memory. The same effect can be obtained by using different access flags for the QPs used to communicate with remote machines.

\section{Overview of \sysname}

\subsection{Architecture}

  \begin{figure*}
    \centering
    \includegraphics[width=\textwidth]{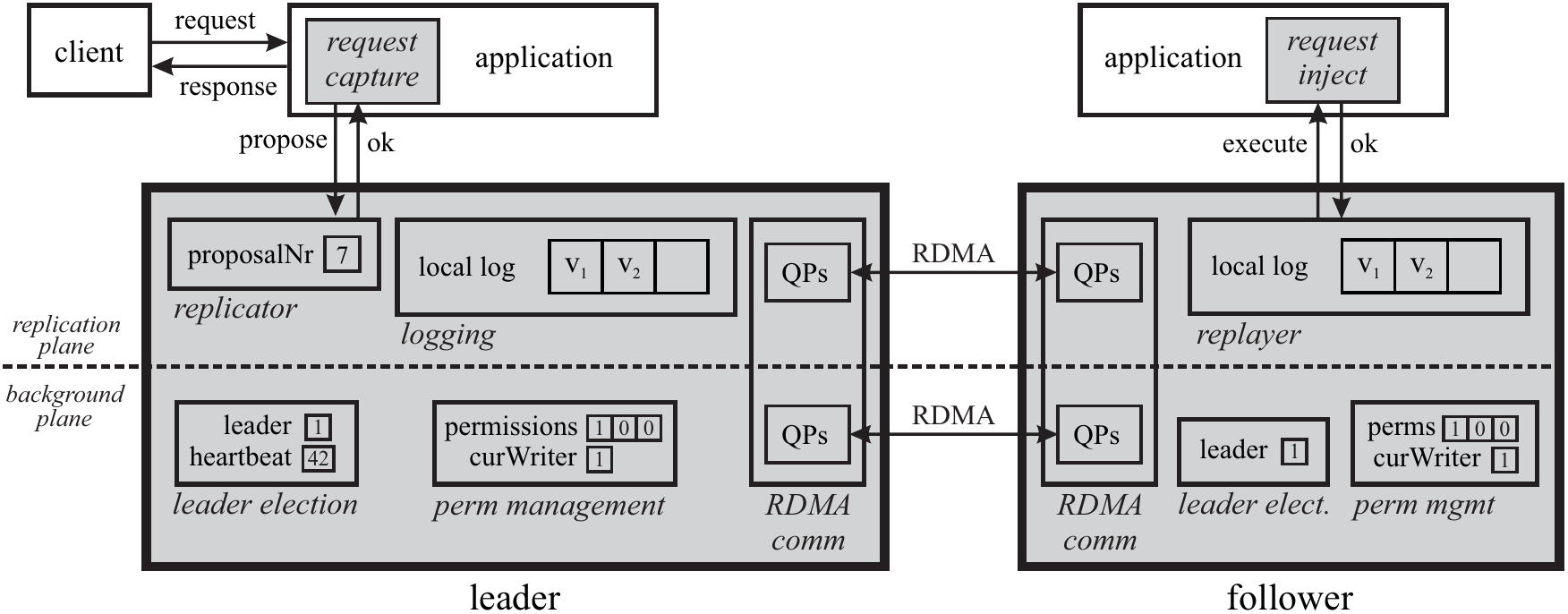}
    \caption{Architecture of \sysname. Grey color shows \sysname components.
    A replica is either a leader or a follower, with different behaviors.
    The leader captures client requests and writes them to the local
    logs of all replicas. Followers replay the log to inject the client
    requests into the application. A leader election component includes
    a heartbeat and the identity of the current leader. A permission
    management component allows a leader to request write permission to the
    local log while revoking the permission from other nodes. \mka{Rachid says: Should we use the Paxos terminology?}
    }
    \label{fig:arch}
\end{figure*}

Figure~\ref{fig:arch} depicts the architecture of \sysname.
At the top, a client sends requests to an application and
  receives a response.
We are not particularly concerned about how the client communicates
  with the application: it can use a network,
  a local pipe, a function call, etc.
We do assume however that this communication is amenable to being
  captured and injected.
That is, there is a mechanism to capture requests from the client
  before they reach the application,
  so we can forward these requests to the replicas; a request
  is an opaque buffer that is not interpreted by \sysname.
Similarly, there is a mechanism to inject requests into the
  app.
Providing such mechanisms requires changing the
  application; however, in our experience, the changes are small
  and non-intrusive.\mka{Rachid says: We do not do that in our applications?}
These mechanisms are standard in any SMR system.

\rmv{  
  \sysname ensures safety irrespective of the number of failures and degree of asynchrony. 
  It ensures liveness provided that a majority of replicas remain alive and communication delays 
  are short.
}

Each replica has an idea of which replica is currently the leader. A replica that considers itself the leader 
assumes that role (left of figure); otherwise, it assumes the role of a follower (right of figure). 
Each replica grants RDMA \emph{write permission} to its log for its current leader and no other replica.
The replicas constantly monitor their current leader to check that it is still active. 
The replicas might not agree on who the current leader is. 
But in \emph{the common case}, all replicas have the same leader, and that leader is active.
%
When that happens, \sysname is simple and efficient.
The leader captures a client request, uses an RDMA Write to append that request to the log of each follower, and then continues
  the application to process the request.
  When the followers detect a new request in their log, they inject
  the request into the application, thereby updating
  the replicas.

The main challenge in the design of SMR protocols is to handle
  leader failures.
Of particular concern is the case when a leader appears failed
  (due to intermittent network delays)
  so another leader takes over, but the original leader
  is still active.

To detect failures in \sysname, the leader periodically increments
  a local counter:  the followers periodically check the counter
  using an RDMA Read.
If the followers do not detect an increment of the counter after a few tries,
a new leader is elected.
  
The new leader revokes a write
  permission by any old leaders, thereby ensuring that
  old leaders cannot interfere with the operation of the new
  leader~\cite{aguilera2019impact}.
The new leader also reconstructs any partial work
  left by prior leaders.
  
Both the leader and the followers are internally divided into two
  major parts: the replication plane and the background plane.
\CRExtra{Roughly, the replication plane plays one of two mutually exclusive roles: the \textit{leader role}, which is responsible for copying
  requests captured by the leader to the followers, or the \textit{follower role}, which replays those requests to update the followers' replicas.}
\CRExtra{The background plane monitors the health of the leader, determines and assigns the leader or follower role to the replication plane,
  and handles permission changes.}
Each plane has its own threads and queue pairs. 
This is in order to improve
  parallelism and provide isolation of performance and functionality.
  More specifically, the following components exist in each of the planes.
  
The replication plane has three components:\mka{Rachid says: Should we use the term ``follower''?}
      \begin{itemize}
         \item {\em Replicator.} This component implements the main protocol to replicate a request from the leader to the followers, by writing the request in the followers' logs using RDMA Write.

        \item {\em Replayer.} This component replays entries from the local log. \CRExtra{This component and the replicator component are mutually exclusive; a replica only has one of these components active, depending on its role in the system.}
        
        \item {\em Logging.} This component stores client requests to be replicated. Each replica has its own local log,
             which may be written remotely by other replicas according to previously granted permissions.
             Replicas also keep a copy of remote logs, which is used by a new leader to reconstruct partial log
             updates by older leaders. 
      \end{itemize}
      
The background plane has two components:
      \begin{itemize}
        \item {\em Leader election.} This component detects failures of leaders and selects other replicas to become leader. \CRExtra{This is what determines the role a replica plays.}
        \item {\em Permission management.} This component grants and revokes write access of local data by remote replicas. It maintains a permissions array, which stores access requests by remote replicas. Basically, a remote replica uses RDMA to store a 1 in this vector to request access.
      \end{itemize}

We describe these planes in more detain in \S\ref{sec:algorithm} and \S\ref{sec:protocol}.

\subsection{RDMA Communication}

Each \node{} has two QPs for each remote \node{}: one
  QP for the \rplane{} and one for the \leplane{}.
The QPs for the \rplane{} share a completion queue,
  while the QPs for the \leplane{} share another completion queue.
The QPs operate in Reliable Connection (RC) mode.

Each \node{} also maintains two MRs, one for each plane.
The MR of the \rplane{} contains the consensus log and the MR of the \leplane{} contains metadata for leader election (\S\ref{sec:leader}) and permission management (\S\ref{sec:permission}). During execution,  \node{s} may change the level of access to their log that they give to each remote \node{}; this is done by changing QP access flags. 
Note that all \node{s} always have remote read and write access permissions to the memory region in the \leplane{} of each \node. 



\section{Replication Plane} \label{sec:algorithm}

The replication plane takes care of execution in the common case, but remains safe during leader changes. This is where we take care to optimize the latency of the common path. We do so by ensuring that, in the replication plane, only a leader replica communicates over the network, whereas all follower replicas are \emph{silent} (i.e., only do local work).

In this section, we discuss algorithmic details related to replication in \sysname{}. 
For pedagogical reasons, we first describe 
in \S\ref{sec:basic-algorithm}
a basic version of the algorithm, \CRExtra{which requires several round-trips to decide.
Later, in \S\ref{sec:extensions}, we discuss how \sysname achieves its single round-trip complexity in the common case, as we present key extensions and optimizations to improve functionality and performance.}
\ifcamera
\else
We give an intuition of why the algorithm works in this section, and we provide the complete correctness argument in the~\hyperref[sec:appendix]{Appendix}.
\fi

\subsection{Basic Algorithm}\label{sec:basic-algorithm}
The leader captures client requests, and calls \emph{propose} to replicate these requests. 
It is simplest to understand our replication algorithm relative to the Paxos
algorithm, which we briefly summarize; for details, we refer the reader to~\cite{paxos}.
%
In Paxos, for each slot of the log, a leader first executes a \emph{prepare phase} where it sends a proposal number to all replicas.\footnote{Paxos uses proposer and acceptor terms; instead, we use leader and replica.}
A replica replies with either nack if it has seen a higher proposal number, or otherwise with the value with the highest proposal number that it has accepted.
After getting a majority of replies, the leader adopts the value with the highest proposal number. 
If it got no values (only acks), it adopts its own proposal value. In the next phase, the \emph{accept phase}, the leader sends its proposal number and adopted value to all replicas. 
A replica acks if it has not received any prepare phase message with a higher proposal number.


In Paxos, replicas actively reply to messages from the leader, but
  in our algorithm, replicas are silent and communicate
  information passively by publishing it to their memory.
Specifically, along with their log, a replica publishes
  a \emph{minProposal} representing the minimum proposal number which it
  can accept. 
The correctness of our algorithm hinges on the leader reading and updating the minProposal number of each follower before updating anything in its log, and on updates on a replica's log happening in slot-order. 

However, this by itself is not enough; Paxos relies on active participation from the followers not only for the data itself, but also to avoid races. Simply publishing the relevant data on each replica is not enough, since two competing leaders could miss each other's updates. This can be avoided if each of the leaders rereads the value after writing it~\cite{gafni2003disk}. However, this requires more communication. To avoid this, we shift the focus from the communication itself to the \emph{prevention} of bad communication. 
A leader $\ell$ maintains a set of \emph{confirmed followers}, which have granted write permission to $\ell$ and revoked write permission from other leaders before $\ell$ begins its operation. This is what prevents races among leaders in \sysname.
%
We describe these mechanisms in more detail below.


\subsubsection*{Log Structure}
The main data structure used by the algorithm is the consensus log kept at each replica (Listing~\ref{alg:log}). 
The log consists of (1) a \textit{minProposal} number, representing the smallest proposal number with which a leader may enter the accept phase on this replica; (2) a \textit{first undecided offset (FUO)}, representing the lowest log index which this replica believes to be undecided; and (3) a sequence of slots---each slot is a $(propNr, value)$ tuple. 
\begin{lstlisting}[label=alg:log, caption=Log Structure,keywords={}]
struct Log {
  minProposal = 0,
  FUO = 0,
  slots[] = (0,@$\bot$@) for all slots 
}
\end{lstlisting}
 
\subsubsection*{Algorithm Description}
Each leader begins its propose call by constructing its \textit{confirmed followers} set (Listing~\ref{alg:replication}, lines~\ref{line:permission-start}--\ref{line:permission-end}). This step is only necessary the first time a new leader invokes propose or immediately after an abort. This step is done by sending permission requests to all replicas and waiting for a majority of acks. When a replica acks, it means that this replica has granted write permission to this leader and revoked it from other replicas. The leader then adds this replica to its confirmed followers set. During execution, if the leader $\ell$ fails to write to one of its confirmed followers, because that follower crashed or gave write access to another leader, $\ell$ aborts and, if it still thinks it is the leader, it calls propose again.

\begin{lstlisting}[label=alg:replication, caption=Basic Replication Algorithm of \sysname,keywords={},firstnumber=last,float]
Propose(myValue):
  done = false
  If I just became leader or I just aborted:
    For every process p in parallel:@\label{line:permission-start}@
      Request permission from p
      If p acks: add p to confirmedFollowers
    Until this has been done for a majority@\label{line:permission-end}@
  While not done:@\label{line:checkDone}@
    Execute Prepare Phase
    Execute Accept Phase
  
Prepare Phase:
  For every process p in confirmedFollowers:
    Read minProposal from p's log @\label{line:readMinProp}@
  Pick a new proposal number, propNum, higher than any minProposal seen so far @\label{line:pickProp}@
  For every process p in confirmedFollowers:
    Write propNum into LOG[p].minProposal @\label{line:writePrepare}@
    Read LOG[p].slots[myFUO] @\label{line:readVals}@
    Abort if any write fails @\label{line:abort-phase1}@
  If all entries read were empty: @\label{line:checkEmpty}@
    value = myValue @\label{line:adoptOwn}@
  Else:
    value = entry value with the largest proposal number of slots read @\label{line:freshestValue}@
  
Accept Phase:
  For every process p in confirmedFollowers:
    Write propNum,value to p in slot myFUO @\label{line:writeAccept}@
    Abort if any write fails @\label{line:abort-phase2}@
  If value == myValue: @\label{line:checkMyValue}@
    done = true @\label{line:setDone}@
  Locally increment myFUO @\label{line:incrementFUO}@
\end{lstlisting}
After establishing its confirmed followers set, the leader invokes the prepare phase. 
To do so, the leader reads the \textit{minProposal} from its confirmed followers (line~\ref{line:readMinProp}) and chooses a proposal number \textit{propNum} which is larger than any that it has read or used before. 
%
%
Then, the leader writes its proposal number into \textit{minProposal} for each of its confirmed followers. Recall that if this write fails at any follower, the leader aborts.
It is safe to overwrite a follower $f$'s \textit{minProposal} in line~\ref{line:writePrepare} because, if that write succeeds, then $\ell$ has not lost its write permission since adding $f$ to its confirmed followers set, meaning no other leader wrote to $f$ since then. 
To complete its prepare phase, the leader reads the relevant log slot of all of its confirmed followers and, as in Paxos, adopts either (a) the value with the highest proposal number, if it read any non-$\bot$ values, or (b) its own initial value, otherwise.


The leader $\ell$ then enters the accept phase, in which it tries to commit its previously adopted value. To do so, $\ell$ writes its adopted value to its confirmed followers. If these writes succeed, then $\ell$ has succeeded in replicating its value. No new value or minProposal number could have been written on any of the confirmed followers in this case, because that would have involved a loss of write permission for $\ell$. Since the confirmed followers set constitutes a majority of the replicas, this means that $\ell$'s replicated value now appears in the same slot at a majority.

Finally, $\ell$ increments its own FUO to denote successfully replicating a value in this new slot. If the replicated value was $\ell$'s own proposed value, then it returns from the \textit{propose} call; otherwise it continues with the prepare phase for the new FUO. 



\subsection{Extensions}\label{sec:extensions}
The basic algorithm described so far is clear and concise, but it also has downsides related to functionality and performance.
We now address these downsides with some extensions, all of which are standard for Paxos-like algorithms; 
\ifcamera
\else
their correctness is discussed in the~\hyperref[sec:appendix]{Appendix}.
\fi

\paragraph{Bringing stragglers up to date.} In the basic algorithm, if a replica $r$ is not included in some leader's confirmed followers set, then its log will lag behind. If $r$ later becomes leader, it can catch up by proposing new values at its current FUO, discovering previously accepted values, and re-committing them. This is correct but inefficient. Even worse, if $r$ never becomes leader, then it will never recover the missing values. We address this problem by introducing an update phase for new leaders. After a replica becomes leader and establishes its confirmed followers set, but before attempting to replicate new values, the new leader (1) brings itself up to date with its highest-FUO confirmed follower (Listing~\ref{alg:catchup}) and (2) brings its followers up to date (Listing~\ref{alg:updateFollowers}). This is done by copying the contents of the more up-to-date log to the less up-to-date log. 
\begin{lstlisting}[label={alg:catchup}, caption={Optimization: Leader Catch Up}, float=ht]
  For every process p in confirmedFollowers
    Read p's FUO
    Abort if any read fails
  F = follower with max FUO
  if F.FUO > myFUO:
    Copy F.LOG[myFUO: F.FUO] into my log
    myFUO = F.FUO
    Abort if the read fails
\end{lstlisting}

\begin{lstlisting}[label={alg:updateFollowers}, caption={Optimization: Update Followers}]
  For every process p in confirmed followers:
    Copy myLog[p.FUO: myFUO] into p.LOG
    p.FUO = myFUO
    Abort if any write fails
\end{lstlisting}

\paragraph{Followers commit in background.} In the basic algorithm, followers do not know when a value is committed and thus cannot replay the requests in the application. This is easily fixed without additional communication. Since a leader will not start replicating in an index $i$ before it knows index $i-1$ to be committed, followers can monitor their local logs and commit all values up to (but excluding) the highest non-empty log index. This is called \emph{commit piggybacking}, since the commit message is folded into the next replicated value.
\CRExtra{As a result, followers replicate but do not commit the $(i{-}1)$-st entry until either the $i$-th entry is proposed by the current leader, or a new leader is elected and brings its followers up to date, whichever happens first.}

\paragraph{Omitting the prepare phase.} Once a leader finds only empty slots at a given index at all of its confirmed followers at line~\ref{line:readVals}, then no higher index may contain an accepted value at any confirmed follower; thus, the leader may omit the prepare phase for higher indexes (until it aborts, after which the prepare phase becomes necessary again). This optimization concerns performance on the common path. With this optimization, the cost of a Propose call becomes a single RDMA write to a majority in the common case.

\paragraph{Growing confirmed followers.} In the \CRExtra{algorithm so far,} the  confirmed followers set remains fixed after the leader initially constructs it. This  implies that processes outside the leader's confirmed followers set will miss updates, even if they are alive and timely, and that the leader will abort even if one of its followers crashes. \CRExtra{To avoid this problem, we extend the algorithm to allow the leader 
to grow its confirmed followers set by briefly waiting for responses from all replicas during its
initial request for permission. 
The leader can also add confirmed followers later, but must bring these replicas up to date (using the mechanism described above in \textit{Bringing stragglers up to date}) before adding them to its set.}
When its confirmed follower set is large, the leader cannot wait for its RDMA reads and writes to complete at all of its confirmed followers before continuing, since we require the algorithm to continue operating despite the failure of a minority of the replicas; instead, the leader waits for just a majority of the replicas to complete. 


\paragraph{Replayer.}

Followers continually monitor the log for new entries. This creates a challenge: how to ensure that the follower does not read an incomplete entry that has not yet been fully written by the leader. 
We adopt a standard approach: we add an extra \textit{canary byte}  at the end of each log entry~\cite{LiuWP04,wang2017apus}. Before issuing an RDMA Write to replicate a log entry, the leader sets the entry's canary byte to a non-zero value. The follower first checks the canary and then the entry contents. In theory, it is possible that the canary gets written 
before the other
contents under RDMA semantics. In practice, however, NICs provide left-to-right
semantics in certain cases (\eg, the memory region is in the same NUMA domain as the NIC),
which ensures that the canary is written last. This assumption is made by other
RDMA systems~\cite{dragojevic2014farm,farm2,kalia2014using,LiuWP04, wang2017apus}.
Alternatively,
  we could store a checksum of the data in the canary, and the follower 
  could read the canary
  and wait for the checksum to match the data.
\section{Background Plane} \label{sec:protocol}

The background plane has two main roles: electing and monitoring the leader, and handling permission change requests. In this section, we describe these mechanisms.

\subsection{Leader Election} \label{sec:leader}

The \textit{leader election component} of the background plane maintains an estimate of the current leader, which it continually updates. The replication plane uses this estimate to determine whether to execute as leader or follower. 

Each replica independently and locally decides who it considers to be leader. We opt for a simple rule: replica $i$ decides that $j$ is leader if $j$ is the replica with the lowest id, among those that $i$ considers to be alive.




To know whether a \node{} has failed, we employ a \emph{pull-score} mechanism, based on a \textit{local heartbeat} counter. A leader election thread continually increments its own counter locally and uses RDMA Reads to read the counters (heartbeats) of other \node{s} and check whether they have been updated. It maintains a \emph{score} for every other \node. If a \node{} has updated its counter since the last time it was read, we increment that \node{'s} score; otherwise, we decrement it. \CR{The score is capped by configurable minimum and maximum values, chosen experimentally to  be $0$ and $15$, respectively.} Once a \node{'s} score drops below a \textit{failure threshold}, we consider it to have failed; if its score goes above a \textit{recovery threshold}, we consider it to be active and timely. To avoid oscillation, we have different \textit{failure} and \textit{recovery} thresholds, \CR{chosen experimentally to be $2$ and $6$,} respectively, so as to avoid false positives.

\paragraph{Large network delays.} 
\CR{Mu employs two timeouts: a small timeout in our detection algorithm (scoring), and a longer timeout built into the RDMA connection mechanism.
The small timeout detects crashes quickly under common failures (process crashes, host crashes) without false positives.
The longer RDMA timeout fires only under larger network delays (connection breaks, counter-read failures).
In theory, the RDMA timeout could use exponential back-off to handle unknown delay bounds.
In practice, however, that is not necessary, since we target datacenters with small delays.
}

\paragraph{Fate sharing.}
\CR{
Because replication and leader election run in independent threads,
the replication thread could fail or be delayed, while the leader election thread remains active and timely.
This scenario is problematic if it occurs on a leader, as
the leader cannot commit new entries, and no other leader can be elected. 
To address thie problem, every $X{=}10000$ iterations, the leader election thread checks the replication thread for activity; if the replication thread is stuck inside a call to \texttt{propose}, the replication thread stops incrementing the local counter, to allow a new leader to be elected.
}

\subsection{Permission Management} \label{sec:permission}

The permission management module is used when changing leaders.
Each replica maintains the invariant that only one replica at a time 
has write permission on its log. 
As explained in Section~\ref{sec:algorithm}, when a leader changes in \sysname, the new leader must request write permission from all the other \node{s}; this is done through a simple RDMA Write to a \textit{permission request array} on the remote side. When a replica $r$ sees a \textit{permission request} from a would-be leader $\ell$, $r$ revokes write access from the current holder, grants write access to $\ell$, and sends an ack to $\ell$.

During the transition phase between leaders, it is possible that several \node{s} think themselves to be leader, and thus the permission request array may contain multiple entries. A permission management thread monitors and handles permission change requests one by one in order of requester id by spinning on the local permission request array.

RDMA provides multiple mechanisms to grant and revoke write access. The first mechanism is to register the consensus log as multiple, completely overlapping RDMA memory regions (MRs), one per remote replica. In order to grant or revoke access from replica $r$, it suffices to re-register the MR corresponding to $r$ with different access flags. The second mechanism is to revoke $r$'s write access by moving $r$'s QP to a non-operational state (e.g., \textit{init}); granting $r$ write access is then done by moving $r$'s QP back to the \textit{ready-to-receive (RTR)} state. The third mechanism is to grant or revoke access from replica $r$ by changing the access flags on $r$'s QP.

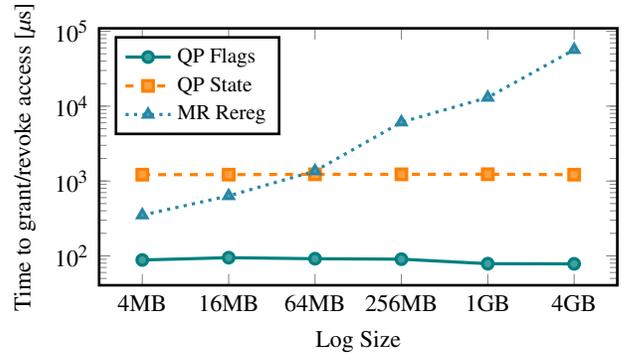
\begin{figure}
\centering
\begin{tikzpicture}
	
    \begin{axis}[
        height=5cm,
        width=\linewidth,
        ytick scale label code/.code={},
        xlabel={Log Size},
        ylabel={Time to grant/revoke access [$\mu$s]},
        xtick=data,
        xticklabels from table={data/permission_change.txt}{size},
        xmode=normal,
        ymode=log,
        legend pos=north west,
        legend style={
            cells={anchor=west}}
    ]
        
    \addplot table[x=x-pos,y=qp_perms] {data/permission_change.txt};
    \addlegendentry{QP Flags};
    
    \addplot table[x=x-pos,y=qp_restart] {data/permission_change.txt};
    \addlegendentry{QP State};
    
    \addplot table[x=x-pos,y=mr_rereg] {data/permission_change.txt};
    \addlegendentry{MR Rereg}

    \end{axis}
	
\end{tikzpicture}
\\

\caption{Performance comparison of different permission switching mechanisms. \textit{QP Flags}: change the access flags on a QP; \textit{QP Restart}: cycle a QP through the \textit{reset}, \textit{init}, \textit{RTR} and \textit{RTS} states; \textit{MR Rereg}: re-register an RDMA MR with different access flags.}
\label{fig:permissions}
\end{figure}
We compare the performance of these three mechanisms in Figure~\ref{fig:permissions}, as a function of the log size (which is the same as the RDMA MR size).
We observe that the time to re-register an RDMA MR grows with the size of the MR, and can reach values close to $100$ms for a log size of $4$GB. On the other hand, the time to change a QPs access flags or cycle it through different states is independent of the MR size, with the former being roughly 10 times faster than the latter. However, changing a QPs access flags while RDMA operations to that QP are in flight sometimes causes the QP to go into an error state. Therefore, in \sysname we use a fast-slow path approach: we first optimistically try to change permissions using the faster QP access flag method and, if that leads to an error, switch to the slower, but robust, QP state method.

\subsection{Log Recycling}
Conceptually, a log is an infinite data structure but in practice
  we need to implement a circular log with limited memory.
This is done as follows.
Each follower has a local \textit{log head} variable, pointing to the first entry not yet executed in its copy of the application. The replayer thread advances the log head each time it executes an entry in the application. Periodically, the leader's \leplane{} reads the log heads of all followers and computes \textit{minHead}, the minimum of all log head pointers read from the followers. Log entries up to the minHead can be reused. Before these entries can be reused, they must be zeroed out to ensure the correct function of the canary byte mechanism.
Thus, the leader zeroes all follower logs after the leader's first undecided offset and before minHead, using an RDMA Write per follower. Note that this means that a new leader must first execute all leader change actions, ensuring that its first undecided offset is higher than all followers' first undecided offsets, before it can recycle entries. To facilitate the implementation, we ensure that the log is never completely full.


\subsection{Adding and Removing Replicas}

\sysname adopts a standard method to add or remove replicas: use
  consensus itself to inform replicas about the change~\cite{paxos}.
More precisely, there is a special log entry that
  indicates that replicas have been removed or added.
Removing replicas is easy: once a replica sees it has been removed,
  it stops executing, while other replicas subsequently
  ignore any communication with it.
Adding replicas is more complicated because it requires
  copying the state of an existing replica into the new one.
To do that, \sysname uses the standard approach of check-pointing state; we do so from one of the followers~\cite{wang2017apus}.
  



\section{Implementation}\label{sec:impl} \label{sec:opt}

\sysname is implemented in 7157 lines of C++17 code (CLOC~\cite{tool-cloc}).
It uses the \emph{ibverbs} library for RDMA over Infiniband.
We implement all features and extensions in sections~\ref{sec:algorithm} and \ref{sec:protocol}, except adding/removing
  replicas and fate sharing.
%
Moreover, we implement some standard RDMA optimizations to reduce latency. 
RDMA Writes and Sends with payloads below a device-specific limit (256 bytes in our setup) are inlined: their payload is written directly to their work request.
We pin threads to cores in the NUMA node of the NIC. 

\CRExtra{Our implementation is modular. We create several modules on top of the \textit{ibverbs} library, which we expose as Conan~\cite{conan} packages. Our modules deal with common practical problems in RDMA-based distributed computing (e.g., writing to all and waiting for a majority, gracefully handling broken RDMA connections etc.). Each abstraction is independently reusable. Our implementation also provides a QP exchange layer, making it straightforward to create, manage, and communicate QP information.}




\section{Evaluation}

Our goal is to evaluate whether \sysname indeed provides viable replication for microsecond computing. We aim to answer the following questions in our evaluation:
\begin{itemize}
    \item What is the replication latency of \sysname? 
    How does it change with payload size and the application being replicated?
    How does \sysname compare to other solutions? 
    \item What is \sysname's fail-over time? 
    \item What is the throughput of \sysname?
\end{itemize}

We evaluate \sysname on a 4-node cluster, the details of which are given in Table~\ref{tab:hwspecs}.
%
All experiments show 3-way replication, which accounts
for most real deployments~\cite{hunt2010zookeeper}.


{\footnotesize
\begin{table}[ht!]
    \centering
    \caption{\CRExtra{Hardware details of machines.}}
	\begin{tabular}{cm{0.64\linewidth}}
	\toprule
\textbf{CPU}		&   2x Intel Xeon E5-2640 v4 @ 2.40GHz \\
\textbf{Memory}	&   2x 128GiB \\
\textbf{NIC}		&   Mellanox Connect-X 4 \\
\textbf{Links}   &   100 Gbps Infiniband \\
\textbf{Switch}  &   Mellanox MSB7700 EDR 100 Gbps  \\
\textbf{OS}      &   Ubuntu 18.04.4 LTS \\
\textbf{Kernel}  &   \texttt{4.15.0-72-generic} \\
\textbf{RDMA Driver} & Mellanox OFED \texttt{4.7-3.2.9.0} \\
	    \bottomrule
	\end{tabular}
    \label{tab:hwspecs}
\end{table}
}

We compare against APUS~\cite{wang2017apus}, DARE~\cite{poke2015dare}, and Hermes~\cite{hermes} where possible. 
\mka{Added sentence about HovercRaft}
The most recent system, HovercRaft~\cite{hovercraft}, also
  provides SMR but its latency at 30--60 microseconds is
  substantially higher than the other systems, so we
  do not consider it further.
For a fair comparison, we disable APUS's persistence to stable storage, since \sysname, DARE, and Hermes all provide only in-memory replication.

We measure time using the POSIX \texttt{clock\_gettime} function, with the \texttt{CLOCK\_MONOTONIC} parameter. 
In our deployment, the resolution and overhead of \texttt{clock\_gettime} is around $16$--$20ns$~\cite{uarch-bench}. In our figures, we show bars labeled with the median latency, with error bars showing 99-percentile and 1-percentile latencies.
These statistics are
computed over 1 million samples with a payload of 64-bytes each, unless otherwise stated.

\paragraph{Applications. }
We use \sysname to replicate several microsecond apps:
  three key-value stores, as well as
  an order matching engine for a financial exchange.

The key-value stores that we replicate with \sysname are Redis~\cite{redis}, Memcached~\cite{memcached}, and HERD~\cite{kalia2014using}. 
For the first two, the client is assumed to be on a different cluster, and connects to the servers over TCP. In contrast, HERD is a microsecond-scale RDMA-based key-value store. We replicate it over RDMA and use it as an example of a microsecond application.
Integration with the three applications requires 183, 228, and 196 additional lines of code, respectively.

The other app is in the context of financial exchanges, in
  which parties unknown to each other submit buy and sell orders of
  stocks, commodities, derivatives, etc.
At the heart of a financial exchange is an order matching 
  engine~\cite{ordermatching}, such as Liquibook~\cite{liquibook},
  which is responsible for matching the buy and sell orders of the parties.
We use \sysname to replicate Liquibook.
Liquibook's inputs are buy and sell orders.
We created an unreplicated client-server version of Liquibook using
   eRPC~\cite{erpc}, and then replicated this system using \sysname.
The eRPC integration and the replication required 611 lines of code in total.


\subsection{Common-case Replication Latency}\label{sec:replication-latency}

We begin by testing the overhead that \sysname introduces in normal execution, when there is no leader failure. For these experiments, we first measure raw replication latency and compare \sysname to other replication systems, as well as to itself under different payloads and attached applications.
%

\paragraph{Effect of Payload and Application on Latency}
We first study \sysname in isolation, to understand its replication latency under different conditions.

We evaluate the raw replication latency of \sysname in two settings: \emph{standalone} and \emph{attached}.
In the standalone setting, \sysname runs just the replication layer with no application and no client; the leader
  simply generates a random payload and invokes
  \texttt{propose()} in a tight loop.
In the attached setting, \sysname is integrated into
  one of a number of applications; the application client
  produces a payload and invokes 
  \texttt{propose()} on the leader.
These settings could impact latency differently,
  \sysname and the application could interfere with each other.

\begin{figure}
    \centering
    \includegraphics[width=\columnwidth]{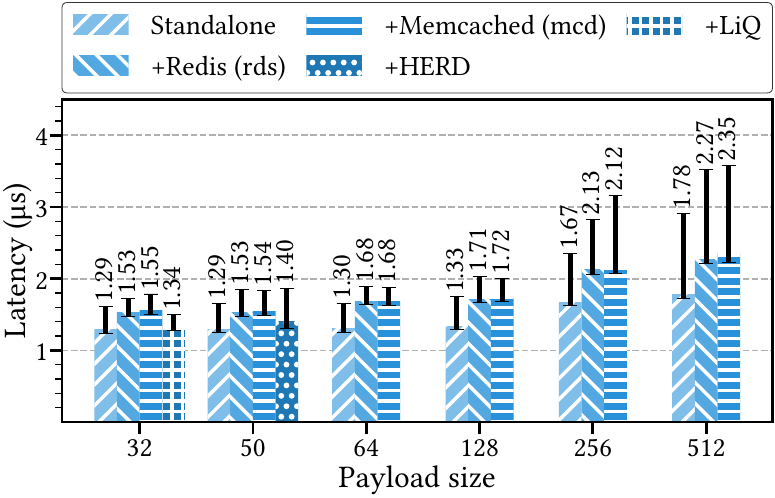}
    \caption{Replication latency of \sysname integrated into
    different applications
    [\memcached (mcd), Liquibook (LiQ), Redis (rds), HERD]
    and payload sizes. Bar height and numerical labels show median latency; error bars show 99-percentile and 1-percentile latencies.}
    \label{fig:usvsus}
\end{figure}

Figure~\ref{fig:usvsus} compares standalone to attached runs as we vary payload size. Liquibook and Herd allow only one payload size (32 and 50 bytes), so they have only one bar each in the graph, while Redis and Memcached have many bars. 


We see that the standalone version slightly outperforms the attached runs, for all tested applications and payload sizes. 
This is due to processor cache effects; in standalone runs, replication state, such as log and queue pairs, are always in cache, and the requests themselves need not be fetched from memory. This is not the case when attaching to an application.
\CRExtra{Additionally, in attached runs, the OS can migrate application threads (even if \sysname's threads are pinned), leading to additional cache effects which can be detrimental to performance.}

\sysname supports two ways of attaching to an application, which
  have different processor cache sharing effects.
The \emph{direct} mode uses the same thread to run both the
  application and the replication, and so they share L1 and L2
  caches.
In contrast, the \emph{handover} method places the application
  thread on a separate core from the replication thread, thus
  avoiding sharing L1 or L2 caches.
Because the application must communicate the request to
  the replication thread, the handover method requires
  a cache coherence miss per replicated request.
This method consistently adds ${\approx}400$ns over the standalone   method. 
For applications with large requests, this overhead might
  be preferable to the one caused by the direct method, where
  replication and application compete for CPU time.
For lighter weight applications, the direct method is preferable.
In our experiments, we measure both methods and show the best method for each application: Liquibook and HERD use the direct method, while Redis and Memcached use the handover method. 

We see that for payloads under 256 bytes, standalone latency   
  remains constant despite increasing payload size. 
This is because we can RDMA-inline requests for these payload
  sizes, so the amount of work needed to send a request remains practically the same.
At a payload of 256 bytes, the NIC must do a DMA itself to fetch the value to be sent, which incurs a gradual increase in overhead
as the payload size increases.
However, we see that \sysname still performs well even at larger payloads quite well; at 512B, the median latency is only $35\%$ higher than the latency of inlined payloads.

\paragraph{Comparing \sysname to Other Replication Systems.}
We now study the replication time of \sysname compared to other
   replication systems, for various applications.
This comparison is not possible for every pair of
  replication system and application, because some
  replication systems are incompatible with certain
  applications.
In particular, APUS works only with socket-based applications
  (Memcached and Redis).
In DARE and Hermes, the replication protocol is bolted onto 
  a key-value store,
  so we cannot attach it to the apps we consider---instead,
  we report their performance with their key-value stores.


\begin{figure}
    \centering
    \includegraphics[width=\columnwidth]{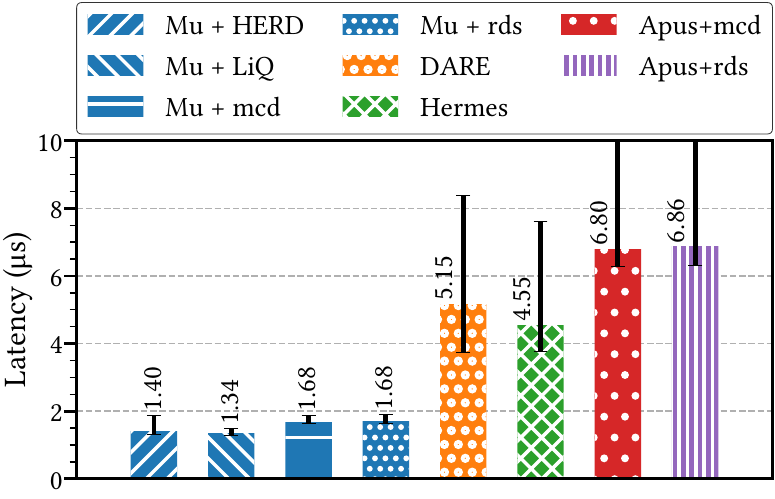}
    \caption{Replication latency of \sysname compared with other
    replication solutions: DARE, Hermes, Apus on memcached (mcd),
    and Apus on Redis (rds). Bar height and numerical labels show median latency; error bars show 99-percentile and 1-percentile latencies.
    }
    \label{fig:replication-usvsthem}
\end{figure}

\begin{figure*}[ht]
    \centering
    \includegraphics[width=0.99\textwidth]{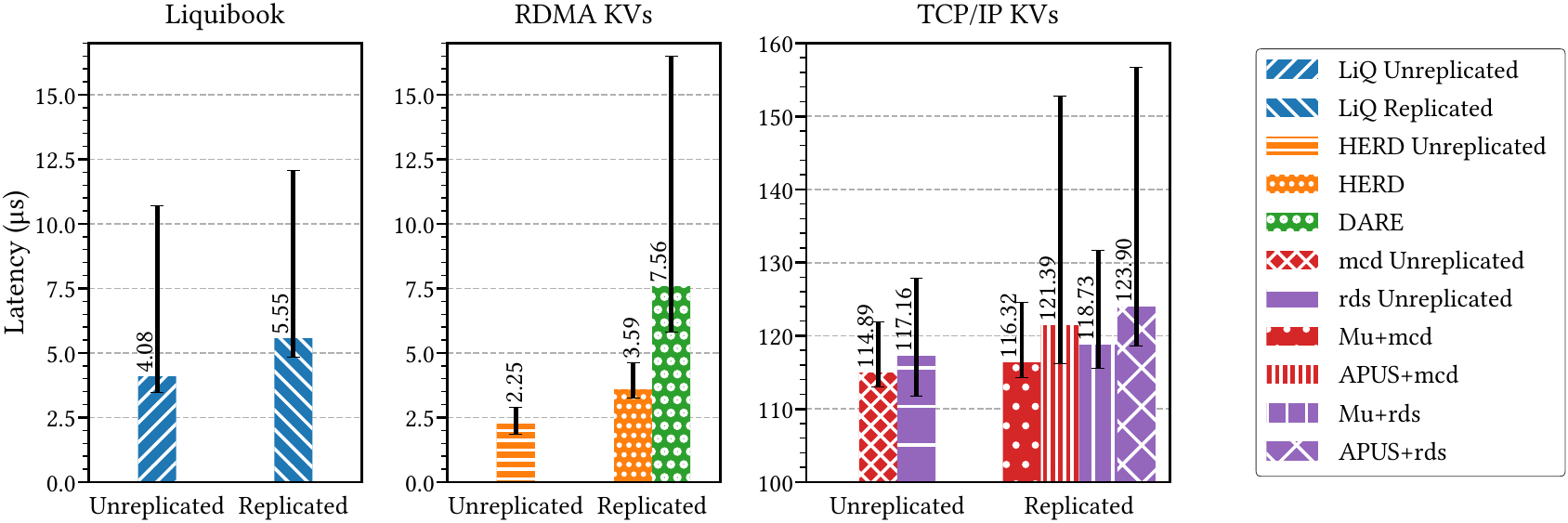}
    
    \caption{End-to-end latencies of applications. 
    The first graph shows a financial exchange app
       (Liquibook) unreplicated and replicated with \sysname. 
    The second graph shows microsecond key-value 
       stores: HERD unreplicated, HERD replicated with \sysname, and DARE.
    The third graph shows traditional key-value stores: Memcached and Redis, unreplicated, as well as replicated with \sysname and APUS. 
    Bar height and numerical labels show median latency; error bars show 99-percentile and 1-percentile latencies.
    }
    
    \label{fig:endtoend}
\end{figure*}

Figure~\ref{fig:replication-usvsthem} shows the replication latencies of these systems.
%
\sysname's median latency outperforms all competitors by at least $2.7\times$, outperforming APUS on the same applications by $4\times$. Furthermore, \sysname has smaller tail variation, with a difference of at most $500$ns between the 1-percentile and 99-percentile latency. In contrast, Hermes and DARE both varied by more than $4\mu s$ across our experiments, with APUS exhibiting 99-percentile executions up to $20\mu s$ slower (cut off in the figure). We attribute this higher variance to two factors: the need to involve the CPU of many replicas in the critical path (Hermes and APUS), and sequentializing several RDMA operations so that their variance aggregates (DARE and APUS). 

\subsection{End-to-End Application Latency} Figure~\ref{fig:endtoend} shows the end-to-end latency of our tested applications, which
includes the latency incurred by the application and by
replication (if enabled).
We show the result in three graphs corresponding to three
  classes of applications.

\CR{In all three graphs, we first focus on the unreplicated latency of these applications, so as to characterize the workload distribution. 
Subsequently, we show the latency of the same applications under replication with \sysname and with competing systems, so as to exhibit the overhead of replication.}

The leftmost graph is for Liquibook.
The left bar is the unreplicated version, and the right
  bar is replicated with \sysname.
We can see that the median latency of Liquibook without replication is $4.08\mu s$, and therefore the overhead of replication is around $35\%$.
There is a large variance in latency, even in the unreplicated
  system.
This variance comes from the client-server
  communication of Liquibook, which is based on eRPC.
This variance changes little with replication. 
The other replication systems cannot replicate Liquibook
  (as noted before, DARE and Hermes 
  are bolted onto their app, and APUS can replicate only socket-based applications).
However, extrapolating their latency 
   from Figure~\ref{fig:replication-usvsthem}, they would add
   unacceptable overheads---over 100\% overhead for the
   best alternative (Hermes).


The middle graph in Figure~\ref{fig:endtoend} shows the client-to-client latency of replicated and unreplicated microsecond-scale key-value stores. The first bars in orange
show HERD unreplicated and HERD replicated with \sysname.
The green bar shows DARE's key-value store with its own
  replication system.
The median unreplicated latency of HERD is $2.25\mu s$, and \sysname adds $1.34\mu s$. While this is a significant overhead ($59\%$ of the original latency), this overhead is lower than any alternative. 
We do not show Hermes in this graph since Hermes does not allow for a separate client, and only generates its requests on the servers themselves. 
HERD replicated with \sysname is the best option for a replicated key-value store, with overall median latency $2\times$ lower than the next best option, and a much lower variance.

\CR{The rightmost graph in Figure~\ref{fig:endtoend} shows the replication of the traditional key-value stores, Memcached and Redis. 
The two leftmost bars show the client-to-client latencies of unreplicated Memcached and Redis, respectively. 
The four rightmost bars show the client-to-client latencies under replication with \sysname and APUS. 
Note that the scale starts at $100\mu s$ to show better precision.}


\CR{\sysname incurs an overhead of around $1.5\mu s$ to replicate these apps, which is about $5\mu s$ faster than replicating with APUS.
For these TCP/IP key-value stores, client-to-client latency under replication with \sysname is around $5\%$ lower than client-to-client latency under replication with APUS. 
With a faster client-to-app network, this difference would be bigger. In either case, \sysname provides fault-tolerant replication with essentially no overhead for these applications.}

\paragraph{Tail latency.} From Figures~\ref{fig:replication-usvsthem} and~\ref{fig:endtoend}, we see that
applications replicated with DARE and APUS
show large tail latencies and a skew
towards lower values (the median latency is closer to the 1-st percentile than the 99-th percentile).
We believe this tail latency occurs
because DARE and APUS must handle several
successive RDMA events on their critical path, where each event is susceptible
to delay, thereby inflating the tail. Because \sysname involves
fewer RDMA events, its tail is smaller.

Figure~\ref{fig:endtoend} shows an even greater tail for the end-to-end latency of replicated applications. Liquibook has a large tail even in its unreplicated version,
which we believe is due to its client-server communication, since the replication of Liquibook with \sysname has a small
tail (Figure~\ref{fig:replication-usvsthem}).
For \memcached{} and \redis{}, additional sources of tail latency are cache effects and thread migration, as discussed in Section~\ref{sec:replication-latency}. This effect is particularly pronounced when replicating with APUS (third panel of Figure~\ref{fig:endtoend}), because the above contributors are compounded.

\subsection{Fail-Over Time}
We now study \sysname's fail-over time. In these experiments, we run the system and subsequently introduce a leader failure. To get a thorough understanding of the fail-over time, we repeatedly introduce leader failures (1000 times) and plot a histogram of the fail-over times we observe. We also time the latency of permission switching, which corresponds to the time to change leaders after a failure is detected. The detection time is the difference between the total fail-over time and the permission switch time.

We inject failures by delaying the leader, thus making it become temporarily unresponsive. This causes other replicas to observe that the leader's heartbeat has stopped changing, and thus detect a failure.
%


\begin{figure}
    \centering
    \includegraphics[width=\columnwidth]{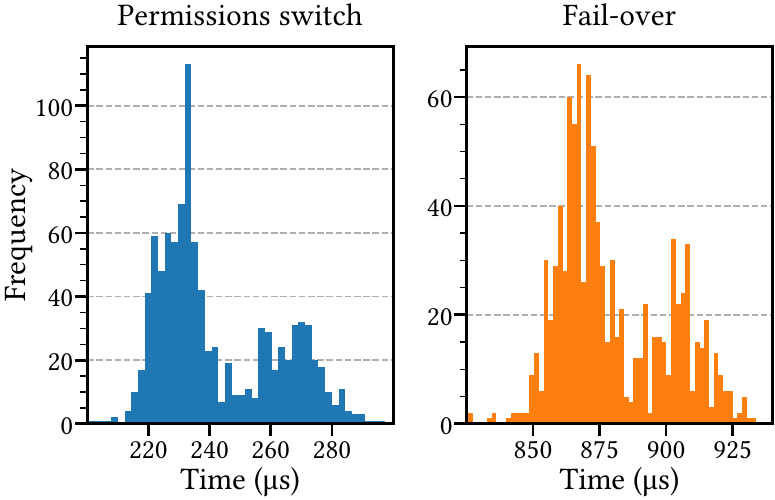}
    
    \caption{Fail-over time distribution.}
    \label{fig:failover}
\end{figure}

Figure~\ref{fig:failover} shows the results. We first note that the total fail-over time is quite low; 
the median fail-over time is $873\mu{s}$ and the 99-percentile fail-over time is $947\mu s$, still below a millisecond. This represents an order of magnitude improvement over the best competitor at
${\approx}10$ ms (HovercRaft~\cite{hovercraft}). 

The time to switch permissions constitutes about $30\%$ of the total fail-over time, with mean latency at $244\mu s$, and 99-percentile at $294\mu s$.
Recall that this measurement in fact encompasses two changes of permission at each replica; one to revoke write permission from the old leader and one to grant it to the new leader. Thus, improvements in the RDMA permission change protocol would be doubly amplified in \sysname's fail-over time.

The rest of the fail-over time is attributed to failure detection (${\approx}600 \mu s$). Although our pull-score mechanism does not rely on network variance, there is still variance
introduced by process scheduling (\eg, in rare cases, the leader process is descheduled by the OS for tens of microseconds)---this is what prevented us from using 
smaller timeouts/scores and it is an area under active investigation for microsecond apps~\cite{muback,shenango,zygos,arachne}.




\subsection{Throughput}
While \sysname optimizes for low latency, in this section we evaluate the throughput of \sysname. 
In our experiment, we run a standalone
microbenchmark (not attached to an application). We increase throughput in two ways: by batching requests together before replicating, and by allowing multiple outstanding requests at a time. In each experiment, we vary the maximum number of outstanding requests allowed at a time, and the batch sizes. 

\begin{figure}
    \centering
    \includegraphics[width=\columnwidth]{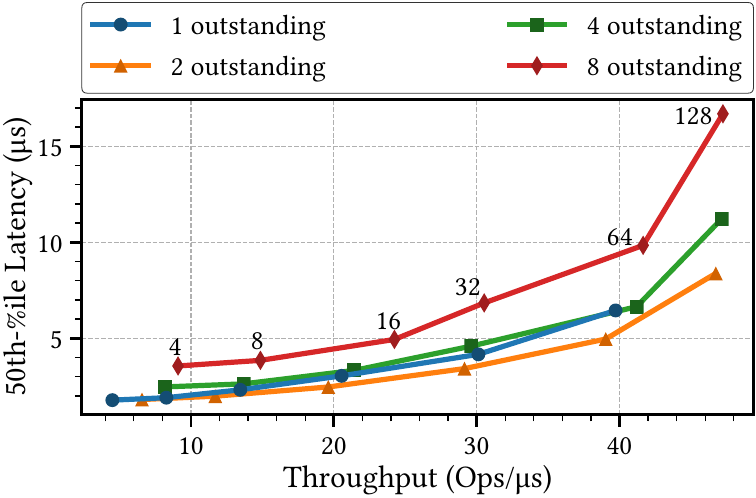}

    \caption{Latency vs throughput. Each line represents a different number of allowed concurrent outstanding requests. Each point on the lines represents a different batch size. Batch size shown as annotation close to each point.}
    \label{fig:throughput}
\end{figure}

Figure~\ref{fig:throughput} shows the results in a
  latency-throughput graph.
Each line represents a different max number of outstanding requests, and each data point represents a different batch size. As before, we use 64-byte requests.

We see that \sysname reaches high throughput with this simple technique. At its highest point, the throughput reaches $47$ Ops/$\mu s$ with a batch size of 128 and 8 concurrent outstanding requests, with per-operation median latency at $17\mu s$. Since the leader is sending requests to two other replicas, this translates to a throughput of $48$Gbps, around half of the NIC bandwidth. 

Latency and throughput both increase as the batch size increases. Median latency is also higher with more concurrent outstanding requests. However, the latency increases slowly, remaining at under $10\mu s$ even with a batch size of $64$ and $8$ outstanding requests.

There is a throughput wall at around $45$ Ops/$\mu s$, with latency rising sharply. This can be traced to the transition between the client requests and the replication protocol at the leader replica. The leader must copy the request it receives into a memory region prepared for its RDMA write. This memory operation becomes a bottleneck. We could optimize throughput further by allowing direct contact between the client and the follower replicas. However, 
%
that may not be useful
  as 
the application itself might need some of the network bandwidth for its own operation, so the replication protocol should not saturate the network.

Increasing the number of outstanding requests while keeping the batch size constant substantially increases throughput at a small latency cost. The advantage of more outstanding requests is largest with two concurrent requests over one. Regardless of batch size, this allows substantially higher throughput at a negligible latency increase: allowing two outstanding requests instead of one increases latency by at most $400ns$ for up to a batch size of 32, and only $1.1\mu s$ at a batch size of 128, while increasing throughput by $20$--$50\%$ depending on batch size. This effect grows less pronounced with higher numbers of outstanding requests.

Similarly, increasing batch size increases throughput with a low latency hit for small batch sizes, but the latency hit grows for larger batches. Notably, using 2 outstanding requests and a batch size of 32 keeps the median latency at only $3.4\mu s$, but achieves throughput of nearly $30$ Ops/$\mu s$. 

\section{Related Work}\label{sec:related}


\paragraph{SMR in General.}

State machine replication is a common technique for building fault-tolerant, 
highly available services~\cite{schneider1990implementing,paxos}.
Many practical SMR protocols have been designed, addressing simplicity~\cite{ongaro2014search,boichat2003deconstructing,hunt2010zookeeper,burrows2006chubby,lampson1996build}, cost~\cite{lamport2006fast,kotla2007zyzzyva}, and harsher failure assumptions~\cite{kotla2007zyzzyva,castro1999practical,gafni2003disk,base}. 
In the original scheme, which we follow, the order of all operations is agreed upon using consensus instances. 
At a high-level, our \sysname protocol resembles 
the classical Paxos algorithm~\cite{paxos}, but there are some important differences.  
In particular, we leverage RDMA's ability to grant and revoke access permissions to ensure 
that two leader replicas cannot both write a value without recognizing each other's presence. 
This allows us to optimize out participation from the follower replicas, leading to better performance. 
Furthermore, these dynamic permissions guide our unique leader changing mechanism.


Several implementations of Multi-Paxos avoid repeating Paxos's prepare phase for every consensus instance, as long as the same leader remains~\cite{pml2007,lamport2001paxos,mazieres2007paxos}. Piggybacking a commit message onto the next replicated request, 
as is done in \sysname, is also used as a latency-hiding mechanism in~\cite{wang2017apus,mazieres2007paxos}.

Aguilera et al.~\cite{aguilera2018passing} suggested the use of local heartbeats in a leader election algorithm designed for a theoretical message-and-memory model, 
in an approach similar to our pull-score mechanism. However, no system has so far implemented such local heartbeats for leader election in RDMA.

Single round-trip replication has been achieved in several previous works using two-sided sends and receives~\cite{keidar2001cost,lamport2006fast,dutta2005fast, hermes,kotla2007zyzzyva}. Theoretical work has shown that single-shot consensus can be achieved in a single one-sided round trip~\cite{aguilera2019impact}. However, \sysname is the first system to put that idea to work and implement one-sided single round trip SMR. 

Alternative reliable replication schemes totally order only non-conflicting operations~\cite{HoltBZPOC16,ClementsKZMK13,gbcast,generalizedconsensus,crdt,curp,hermes}. 
These schemes require opening the service being replicated to identify which operations commute. 
In contrast, we designed \sysname assuming the replicated service is a black box. 
If desired, several parallel instances of \sysname could be used to replicate concurrent operations that commute.
This could be used to increase throughput in specific applications.

It is also important to notice that we consider ``crash'' failures. 
In particular, we assume nodes cannot behave in a Byzantine manner~\cite{kotla2007zyzzyva,ClementWADM09,castro1999practical}.

\paragraph{Improving the Stack Underlying SMR.}

While we propose a new SMR algorithm adapted to RDMA in order to optimize latency, 
other systems keep a classical algorithm but improve the underlying communication
  stack~\cite{erpc,socksdirect}.
With this approach, somewhat orthogonal to ours, the best reported
  replication latency is 5.5 $\mu s$~\cite{erpc}, almost $5\times$ slower than \sysname.
  HovercRaft~\cite{hovercraft} shifts the SMR from the application layer to the transport layer to avoid IO and CPU bottlenecks on the leader replica. However, their request latency is more than an order of magnitude more than that of \sysname, and they do not optimize fail-over time.
  
Some SMR systems leverage
  recent technologies such as programmable switches and
  NICs~\cite{li2016just,jin2018netchain,istvan2016consensus,ipipe}.
However, programmable networks are not as widely
  available
  as RDMA, which has been commoditized with technologies such as RoCE and iWARP.




\paragraph{Other RDMA Applications.}
More generally, RDMA has recently been the focus of many data center system designs, including key-value stores~\cite{kalia2014using,dragojevic2014farm} and transactions~\cite{wei2015fast,kalia2016fasst}. 
Kalia et al. provide guidelines on the best ways to use RDMA to enhance performance~\cite{kalia2016design}. 
Many of their suggested optimizations are employed by \sysname.
Kalia et al. also advocate the use of two-sided RDMA verbs (Sends/Receives) instead of RDMA Reads in situations in which a single RDMA Read might not suffice. 
However, this does not apply to \sysname, since we know a priori which memory location should be read, and we rarely have to follow up with another read.

\paragraph{Failure detection.}
Failure detection is typically done using timeouts.
Conventional wisdom is that timeouts must be large,
  in the seconds~\cite{falcon},
  though some systems report timeouts as low
  as 10 milliseconds~\cite{hovercraft}.
It is possible to improve detection time
  using inside information~\cite{falcon,pigeon} or
  fine-grained reporting~\cite{albatross},
  which requires changes to apps and/or the infrastructure.
This is orthogonal to our score-based mechanism
  and could be used to further improve \sysname.

\subsection*{Similar RDMA-based Algorithms}
A few SMR systems have recently been designed for RDMA~\cite{poke2015dare,wang2017apus,derecho}, but used RDMA differently from \sysname.

\bigskip
\noindent \textbf{DARE}~\cite{poke2015dare} is the first RDMA-based SMR system. Similarly to \sysname, 
DARE uses only one-sided RDMA verbs executed by the leader to replicate the log in normal execution, \CR{and makes use of permissions when changing leaders.} 
\CR{However, unlike Mu, DARE requires updating the tail pointer of each replica's log in a separate RDMA Write from the one that copies over the new value, which leads to more round-trips for replication.}
DARE's use of permissions does not
  lead to a light-weight mechanism to
  block concurrent leaders, as in \sysname. DARE has a heavier leader election protocol than \sysname's, similar to that of RAFT, in which care is taken to ensure that at most one process considers itself 
leader at any point in time.

\bigskip
\noindent \textbf{APUS}~\cite{wang2017apus} improves upon DARE's throughput. 
However, APUS requires active participation from the follower \node{s} during the replication protocol, resulting in higher latencies. \CR{Thus, it does not achieve the one-sided common-case communication of \sysname. Similarly to DARE and \sysname, APUS uses transitions through queue pair states to allow or deny RDMA access. However, like DARE, it does not use this mechanism to achieve a single one-sided communication round.}

\bigskip
\noindent \textbf{Derecho}~\cite{derecho} provides durable and non-durable SMR, by combining a
  data movement protocol (SMC or RDMC) with a shared-state
  table primitive (SST) for determining when it is safe to deliver messages.
This design yields high throughput but also high latency: a minimum of
  10$\mu s$ for non-durable SMR~\cite[Figure 12(b)]{derecho} and more for durable SMR.
This latency results from a node delaying the delivery of a message
  until all nodes have confirmed its receipt using the SST, which
  takes additional RDMA communication steps compared to \sysname.
  It would be interesting to explore how \sysname's protocol could
  improve Derecho.
  
\bigskip
\noindent \textbf{Aguilera et al}~\cite{aguilera2019impact} \CR{present a one-shot consensus algorithm based on RDMA that solves consensus in a single one-sided communication round in the common case. They model RDMA's one-sided verbs as shared memory primitives which operate only if granted appropriate permissions. Their one-round communication complexity relies on changing permissions, an idea we use in \sysname.
 While that work focuses on a theoretical construction, \sysname is a fully fledged SMR system that needs
 many other mechanisms, such as
 logging, managing state, coordinating instances, recycling instances, handling clients, and permission management. Because these mechanisms are non-trivial, \sysname requires its own proof of correctness\ifcamera\else~(see~\hyperref[sec:appendix]{Appendix})\fi. 
 \sysname also provides an implementation and experimental evaluation not found in~\cite{aguilera2019impact}.}

\section{Conclusion}\label{sec:discussion}


Computers have progressed from batch-processing systems that operate at the time scale of minutes, 
to progressively lower latencies in the seconds, then milliseconds, and now we are in the microsecond revolution.
Work has already started in this space at various
layers of the computing stack. Our contribution fits in this context, by providing
generic microsecond replication for microsecond apps.

\sysname is a state machine replication system that can replicate microsecond applications with little overhead. 
This involved two goals: achieving low latency on the common path,
and minimizing fail-over time to maintain high availability.
To reach these goals, \sysname relies on 
(a) RDMA permissions to replicate a request with a single one-sided operation, as well as
(b) a failure detection mechanism that does not
incur false positives due to common network delays---a property that
permits \sysname to use aggressively small timeout values.

\bibliographystyle{plain}
\bibliography{bib/biblio}

\ifcamera
\else
\newpage
\appendix
\section{Appendix: Proof of Correctness}\label{sec:appendix}
\subsection{Pseudocode of the Basic Version}

\begin{lstlisting}
Propose(myValue):
    done = false
    If I just became leader or I just aborted
        For every process p in parallel:@\label{line:app-permission-start}@
            Request permission from p
            If p acks, add p to confirmedFollowers
        Until this has been done for a majority of processes@\label{line:app-permission-end}@
    While not done:@\label{line:app-checkDone}@
        Execute Prepare Phase
        Execute Accept Phase
\end{lstlisting}

\begin{lstlisting}[firstnumber=last]
struct Log {
    log[] = @$\bot$@ for all slots
    minProposal = 0
    FUO = 0    }
    
Prepare Phase:
    Pick a new proposal number, propNum, that is higher than any seen so far @\label{line:app-pickProp}@
    For every process p in confirmedFollowers:
        Read minProposal from p's log @\label{line:app-readMinProp}@
        Abort if any read fails @\label{line:app-abort-phase0}@
    If propNum < some minProposal read, abort @\label{line:app-abortProposal}@ 
    For every process p in confirmedFollowers:
        Write propNum into LOG[p].minProposal @\label{line:app-writePrepare}@
        Read LOG[p].slots[myFUO] @\label{line:app-readVals}@
        Abort if any write or read fails @\label{line:app-abort-phase1}@
    if all entries read were empty: @\label{line:app-checkEmpty}@
        value = myValue @\label{line:app-adoptOwn}@
    else:
        value = entry value with the largest proposal number of slots read @\label{line:app-freshestValue}@
    
Accept Phase:
    For every process p in confirmedFollowers:
        Write value,propNum to p in slot myFUO @\label{line:app-writeAccept}@
        Abort if any write fails @\label{line:app-abort-phase2}@
    If value == myValue: @\label{line:app-checkMyValue}@
        done = true @\label{line:app-setDone}@
    Locally increment myFUO @\label{line:app-incrementFUO}@
\end{lstlisting}

Note that write permission can only be granted at most once per request; it is impossible to send a single permission request, be granted permission, lose permission and then regain it without issuing a new permission request. This is the way that permission requests work in our implementation, and is key for the correctness argument to go through; in particular, it is important that a leader cannot lose permission between two of its writes to the same follower without being aware that it lost permission.
\subsection{Definitions}
\begin{definition}[Quorum]
A \textit{quorum} is any set that contains at least a majority of the processes.
\end{definition}

\begin{definition}[Decided Value]
We say that a value $v$ is \textit{decided} at index $i$ if there exists a quorum $Q$ such that for every process $p \in Q$, $p$'s log contains $v$ at index $i$.
\end{definition}

\begin{definition}[Committed Value]
We say that a value $v$ is \textit{committed} at process $p$ at index $i$ if $p$'s log contains $v$ at index $i$, such that $i$ is less than $p$'s FUO.
\end{definition}

\subsection{Invariants}

\subsubsection{Preliminary}

\begin{invariant}[Committed implies decided] \label{inv:commit}
If a value $v$ is committed at some process $p$ at index $i$, then $v$ is decided at index $i$.
\end{invariant}
\begin{proof}
Assume $v$ is committed at some process $p$ at index $i$. Then $p$ must have incremented its FUO past $i$ at line~\ref{line:app-incrementFUO}, therefore $p$ must have written $v$ at a majority at line~\ref{line:app-writeAccept}.
\end{proof}


\begin{invariant}[Values are never erased]\label{inv:values-not-erased}
If a log slot contains a value at time $t$, that log slot will always contain some value after time $t$.
\end{invariant}
\begin{proof}
By construction of the algorithm, values are never erased (note: values can be overwritten, but only with a non-$\bot$ value).
\end{proof}

\subsubsection{Validity}
\begin{invariant}\label{inv:pre-validity}
If a log slot contains a value $v\neq \bot$, then $v$ is the input of some process.
\end{invariant}
\begin{proof}
Assume the contrary and let $t$ be the earliest time when some log slot (call it $L$) contained a non-input value (call it $v$). In order for $L$ to contain $v$, some process $p$ must have written $v$ into $L$ at line~\ref{line:app-writeAccept}. Thus, either $v$ was the input value of $p$ (which would lead to a contradiction), or $p$ adopted $v$ at line~\ref{line:app-freshestValue}, after reading it from some log slot $L'$ at line~\ref{line:app-readVals}. Thus, $L'$ must have contained $v$ earlier than $t$, a contradiction of our choice of $t$. 
\end{proof}

\begin{theorem}[Validity]\label{inv:validity}
If a value $v$ is committed at some process, then $v$ was the input value of some process.
\end{theorem}
\begin{proof}
Follows immediately from Invariant~\ref{inv:pre-validity} and the definition of being committed.
\end{proof}

\subsubsection{Agreement}

\begin{invariant}[Solo detection]\label{inv:solo-detection}
If a process $p$ writes to a process $q$ in line~\ref{line:app-writePrepare} or in line~\ref{line:app-writeAccept}, then no other process $r$ wrote to $q$ since $p$ added $q$ to its confirmed followers set.
\end{invariant}
\begin{proof}
Assume the contrary: $p$ added $q$ to its confirmed followers set at time $t_0$ and wrote to $q$ at time $t_2 > t_0$; $r \ne p$ wrote to $q$ at time $t_1, t_0 < t_1 < t_2$. Then:
\begin{enumerate}
    \item $r$ had write permission on $q$ at $t_1$.
    \item $p$ had write permission on $q$ at $t_2$.
    \item (From (1) and (2)) $p$ must have obtained write permission on $q$ between $t_1$ and $t_2$. But this is impossible, since $p$ added $q$ to its confirmed followers set at $t_0 < t_1$ and thus $p$ must have obtained permission on $q$ before $t_0$. By the algorithm, $p$ did not request permission on $q$ again since obtaining it, and by the way permission requests work, permission is granted at most once per request. We have reached a contradiction. \qedhere
\end{enumerate}
\end{proof}

\begin{invariant}\label{inv:pre-agreement}
If some process $p_1$ successfully writes value $v_1$ and proposal number $b_1$ to its confirmed followers in slot $i$ at line~\ref{line:app-writeAccept}, then any process $p_2$ entering the accept phase with proposal number $b_2 > b_1$ for slot $i$ will do so with value $v_1$.
\end{invariant}
\begin{proof}
Assume the contrary: some process enters the accept phase for slot $i$ with a proposal number larger than $b_1$, with a value $v_2 \neq v_1$. Let $p_2$ be the first such process to enter the accept phase. 

Let $C_1$ (resp. $C_2$) be the confirmed followers set of $p_1$ (resp. $p_2$). Since $C_1$ and $C_2$ are both quorums, they must intersect in at least one process, call it $q$. Since $q$ is in the confirmed followers set of both $p_1$ and $p_2$, both must have read its minProposal (line~\ref{line:app-readMinProp}), written its minProposal with their own proposal value (line~\ref{line:app-writePrepare}) and read its $i$th log slot (line~\ref{line:app-readVals}). Furthermore, $p_1$ must have written its new value into that slot (line~\ref{line:app-writeAccept}).
Note that since $p_1$ successfully wrote value $v_1$ on $q$, by Invariant~\ref{inv:solo-detection}, $p_2$ could not have written on $q$ between the time at which $p_1$ obtained its permission on it and the time of $p_1$'s write on $q$'s $i$th slot. Thus, $p_2$ either executed both of its writes on $q$ before $p_1$ obtained permissions on $q$, or after $p_1$ wrote its value in $q$'s $i$th slot. If $p_2$ executed its writes before $p_1$, then $p_1$ must have seen $p_2$'s proposal number when reading $q$'s minProposal in line~\ref{line:app-readMinProp} (since $p_1$ obtains permissions before executing this line). Thus, $p_1$ would have aborted its attempt and chosen a higher proposal number, contradicting the assumption that $b_1 < b_2$.

Thus, $p_2$ must have executed its first write  on $q$ after $p_1$ executed its write of $v_1$ in $q$'s log. Since $p_2$'s read of $q$'s slot happens after its first write (in line~\ref{line:app-readVals}), this read must have happened after $p_1$'s write, and therefore $p_2$ saw $v_1, b_1$ in $q$'s $i$th slot. By assumption, $p_2$ did not adopt $v_1$. By line~\ref{line:app-freshestValue}, this means $p_2$ read $v_2$ with a higher proposal number than $b_1$ from some other process in $C_2$. This contradicts the assumption that $p_2$ was the first process to enter the accept phase with a value other than $v_1$ and a proposal number higher than $b_1$. The figure below illustrates the timings of events in the execution.
\end{proof}

\begin{theorem}[Agreement]\label{inv:agreement}
If $v_1$ is committed at $p_1$ at index $i$ and $v_2$ is committed at $p_2$ at index $i$, then $v_1 = v_2$.
\end{theorem}
\begin{proof}
In order for $v_1$ (resp. $v_2$) to be committed at $p_1$ (resp. $p_2$) at index $i$, $p_1$ (resp. $p_2$) must have incremented its FUO past $i$ and thus
must have successfully written $v_1$ (resp. $v_2$) to its confirmed follower set at line~\ref{line:app-writeAccept}. Let $b_1$ (resp. $b_2$) be the proposal number $p_1$ (resp. $p_2$) used at line~\ref{line:app-writeAccept}. Assume without loss of generality that $b_1 < b_2$. Then, by Invariant~\ref{inv:pre-agreement}, $p_2$ must have entered its accept phase with value $v_1$ and thus must have written $v_1$ to its confirmed followers at line~\ref{line:app-writeAccept}. Therefore, $v_1 = v_2$.
\end{proof}

\subsubsection{Termination}

\begin{invariant}[Termination implies commitment.]
If a process $p$ calls propose with value $v$ and returns from the propose call, then $v$ is committed at $p$.
\end{invariant}
\begin{proof}
Follows from the algorithm: $p$ returns from the propose call only after it sees $done$ to be $true$ at line~\ref{line:app-checkDone}; for this to happen, $p$ must set $done$ to $true$ at line~\ref{line:app-setDone} and increment its FUO at line~\ref{line:app-incrementFUO}. In order for $p$ to set $done$ to $true$, $p$ must have successfully written some value $val$ to its confirmed follower set at line~\ref{line:app-writeAccept} and $val$ must be equal to $v$ (check at line~\ref{line:app-checkMyValue}). Thus, when $p$ increments its FUO at line~\ref{line:app-incrementFUO}, $v$ becomes committed at $p$.
\end{proof}

\begin{invariant}[Weak Termination]\label{inv:weak-termination}
If a correct process $p$ invokes Propose and does not abort, then $p$ eventually returns from the call.
\end{invariant}
\begin{proof}
The algorithm does not have any blocking steps or goto statements, and has only one unbounded loop at line~\ref{line:app-checkDone}. Thus, we show that $p$ will eventually exit the loop at line~\ref{line:app-checkDone}.

Let $t$ be the time immediately after $p$ finishes constructing its confirmed followers set (lines~\ref{line:app-permission-start}--\ref{line:app-permission-end}). Let $i$ be the highest index such that one of $p$'s confirmed followers contains a value in its log at index $i$ at time $t$. Given that $p$ does not abort, it must be that $p$ does not lose write permission on any of its confirmed followers and thus has write permission on a quorum for the duration of its call. Thus, after time $t$ and until the end of $p$'s call, no process is able to write any new value at any of $p$'s confirmed followers [$\ast$].

Since $p$ never aborts, it will repeatedly execute the accept phase and increment its FUO at line~\ref{line:app-incrementFUO} until $p$'s FUO is larger than $i$. During its following prepare phase, $p$ will find all slots to be empty (due to [$\ast$]) and adopt its own value $v$ at line~\ref{line:app-adoptOwn}. Since $p$ does not abort, $p$ must succeed in writing $v$ to its confirmed followers at line~\ref{line:app-writeAccept} and set $done$ to $true$ in line~\ref{line:app-setDone}. Thus, $p$ eventually exits the loop at line~\ref{line:app-checkDone} and returns.
\end{proof}

\begin{theorem}[Termination] If eventually there is a unique non-faulty leader, then eventually every Propose call returns.
\end{theorem}
\begin{proof}
We show that eventually $p$ does not abort from any Propose call and thus, by Invariant~\ref{inv:weak-termination}, eventually $p$ returns from every Propose call.

Consider a time $t$ such that (1) no processes crash after $t$ and (2) a unique process $p$ considers itself leader forever after $t$. 

Furthermore, by Invariant~\ref{inv:weak-termination}, by some time $t' > t$ all correct processes will return or abort from any Propose call they started before $t$; no process apart from $p$ will call Propose again after $t'$ since $p$ is the unique leader.

Thus, in any propose call $p$ starts after $t'$, $p$ will obtain permission from a quorum in lines~\ref{line:app-permission-start}--\ref{line:app-permission-end} and will never lose any permissions (since no other process is requesting permissions). Thus, all of $p$'s reads and writes will succeed, so $p$ will not abort at lines~\ref{line:app-abort-phase0}, \ref{line:app-abort-phase1}, or \ref{line:app-abort-phase2}.

Furthermore, since no process invokes Propose after $t'$, the minProposals of $p$ confirmed followers do not change after this time. Thus, by repeatedly increasing its minProposal at line~\ref{line:app-pickProp}, $p$ will eventually have the highest proposal number among its confirmed followers, so $p$ will not abort at line~\ref{line:app-abortProposal}. 

Therefore, by Invariant~\ref{inv:weak-termination}, $p$ will eventually return from every Propose call.
\end{proof}

We consider a notion of eventual synchrony, whereby after some unknown global stabilization time, all processes become timely. If this is the case, then Mu's leader election mechanism ensures that eventually, a single correct leader is elected forever. This leader is the replica with the lowest id that did not crash: after the global stabilization point, this replica would be timely, and therefore would not miss a heartbeat. All other replicas would see its heartbeats increasing forever, and elect it as their leader. This guarantees that our algorithm terminates under this eventual synchrony condition.
\subsection{Optimizations \& Additions}\label{sec:optimizations}
\subsubsection{New Leader Catch-Up}
In the basic version of the algorithm described so far, it is possible for a new leader to miss decided entries from its log (e.g., if the new leader was not part of the previous leader's confirmed followers). The new leader can only catch up by attempting to propose new values at its current FUO, discovering previously accepted values, and re-committing them. This is correct but inefficient.

We describe an extension that allows a new leader $\ell$ to catch up faster: after constructing its confirmed followers set (lines~\ref{line:app-permission-start}--\ref{line:app-permission-end}), $\ell$ can read the FUO of each of its confirmed followers, determine the follower $f$ with the highest FUO, and bring its own log and FUO up to date with $f$. This is described in the pseudocode below:

\begin{lstlisting}[label={alg:app-catchup}, caption={Optimization: Leader Catch Up}]
    For every process p in confirmedFollowers
        Read p's FUO
        Abort if any read fails
    F = follower with max FUO
    if F.FUO > my_FUO:
        Copy F.LOG[my_FUO: F.FUO] into my log
        myFUO = F.FUO
        Abort if any read fails
\end{lstlisting}

We defer our correctness argument for this extension to Section~\ref{sec:update-followers}.



\subsubsection{Update Followers}\label{sec:update-followers}

While the previous extension allows a new leader to catch up in case it does not have the latest committed values, followers' logs may still be left behind (e.g., for those followers that were not part of the leader's confirmed followers).

As is standard for practical Paxos implementations, we describe a mechanism for followers' logs to be updated so that they contain all committed entries that the leader is aware of. After a new leader $\ell$ updates its own log as described in Algorithm~\ref{alg:app-catchup}, it also updates its confirmed followers' logs and FUOs:

\begin{lstlisting}[label={alg:app-updateFollowers}, caption={Optimization: Update Followers}]
    For every process p in confirmed followers:
        Copy myLog[p.FUO: my_FUO] into p.LOG
        p.FUO = my_FUO
        Abort if any write fails
\end{lstlisting}

We now argue the correctness of the update mechanisms in this and the preceding subsections. These approaches clearly do not violate termination. We now show that they preserve agreement and validity.

\begin{proof}[Validity.]
We extend the proof of Invariant~\ref{inv:pre-validity} to also cover Algorithms~\ref{alg:app-catchup} and \ref{alg:app-updateFollowers}; the proof of Theorem~\ref{inv:validity} remains unchanged. 

Assume by contradiction that some log slot $L$ does not satisfy Invariant~\ref{inv:pre-validity}. Without loss of generality, assume that $L$ is the first log slot in the execution which stops satisfying Invariant~\ref{inv:pre-validity}. In order for $L$ to contain $v$, either (i) some process $q$ wrote $v$ into $L$ at line~\ref{line:app-writeAccept}, or (ii) $v$ was copied into $L$ using Algorithm~\ref{alg:app-catchup} or \ref{alg:app-updateFollowers}. In case (i), either $v$ was $q$'s input value (a contradiction), or $q$ adopted $v$ at line~\ref{line:app-freshestValue} after reading it from some log slot $L'\ne L$. In this case, $L'$ must have contained $v$ before $L$ did, a contradiction of our choice of $L$. In case (ii), some log slot $L''$ must have contained $v$ before $L$ did, again a contradiction.
\end{proof}

\begin{proof}[Agreement.]
We extend the proof of~\ref{inv:agreement} to also cover Algorithms~\ref{alg:app-catchup} and \ref{alg:app-updateFollowers}. Let $t$ be the earliest time when agreement is broken; i.e., $t$ is the earliest time such that, by time $t$, some process $p_1$ has committed $v_1$ at $i$ and some process $p_2$ has committed $v_2 \ne v_1$ at $i$. We can assume without loss of generality that $p_1$ commits $v_1$ at $t_1 = t$ and $p_2$ commits $v_2$ at $t_2 < t_1$. We now consider three cases:

\begin{enumerate}
    \item Both $p_1$ and $p_2$ commit normally by incrementing their FUO at line~\ref{line:app-incrementFUO}. Then the proof of~\ref{inv:agreement} applies to $p_1$ and $p_2$.
    \item $p_1$ commits normally by incrementing its FUO at line~\ref{line:app-incrementFUO}, while $p_2$ commits with Algorithm~\ref{alg:app-catchup} or \ref{alg:app-updateFollowers}. Then some process $p_3$ must have committed $v_2$ normally at line~\ref{line:app-incrementFUO} and the proof of ~\ref{inv:agreement} applies to $p_1$ and $p_3$.
    \item $p_1$ commits $v_1$ using Algorithm~\ref{alg:app-catchup} or \ref{alg:app-updateFollowers}. Then $v_1$ was copied to $p_1$'s log from some other process $p_3$'s log, where $v_1$ had already been committed. But then, agreement must have been broken earlier than $t$ ($v_1$ committed at $p_3$, $v_2$ committed at $p_2$), a contradiction.
\end{enumerate}
\end{proof}


\subsubsection{Followers Update Their Own FUO}
In the algorithm and optimizations presented so far, the only way for the FUO of a process $p$ to be updated is by the leader; either by $p$ being the leader and updating its own FUO, or by $p$ being the follower of some leader that executes Algorithm~\ref{alg:app-updateFollowers}. However, in the steady state, when the leader doesn't change, it would be ideal for a follower to be able to update its own FUO. This is especially important in practice for SMR, where each follower should be applying committed entries to its local state machine. Thus, knowing which entries are committed as soon as possible is crucial. For this purpose, we introduce another simple optimization, whereby a follower updates its own FUO to $i$ if it has a non-empty entry in some slot $j \geq i$ and all slots $k < i$ are populated.

\begin{lstlisting}[label={alg:FUO}, caption={Optimization: Followers Update Their Own FUO}]
    if LOG[i] @$\neq \bot$@ && my_FUO == i-1
        my_FUO = i
\end{lstlisting}

Note that this optimization doesn't write any new values on any slot in the log, and therefore, cannot break Validity. Furthermore, since it does not introduce any waiting, it cannot break termination.
We now prove that this doesn't break Agreement.

\begin{proof}[Agreement.]
Assume by contradiction that executing Algorithm~\ref{alg:FUO} can break agreement. Let $p$ be the first process whose execution of Algorithm~\ref{alg:FUO} breaks agreement, and let $t$ be the time at which it changes its FUO to $i$, thereby breaking agreement.

It must be the case that $p$ has all slots up to and including $i$ populated in its log. Furthermore, since $t$ is the first time at which disagreement happens, and $p$'s FUO was at $i-1$ before $t$, it must be the case that for all values in slots $1$ to $i-2$ of $p$'s log, if any other process $p'$ also has those slots committed, then it has the same values as $p$ in those slots. 
Let $p$'s value at slot $i-1$ be $v$. Let $\ell_1$ be the leader that populated slot $i-1$ for $p$, and let $\ell_2$ be the leader the populated slot $i$ for $p$. If $\ell_1 = \ell_2$, then $p$'s entry at $i-1$ must be committed at $\ell_1$ before time $t$, since otherwise $\ell_1$ would not have started replicating entry $i$. So, if at time $t$, some process $q$ has a committed value $v'$ in slot $i-1$ where $v' \neq v$, then this would have violated agreement with $\ell_1$ before $t$, contradicting the assumption that $t$ is the earliest time at which agreement is broken.

Now consider the case where $\ell_1 \neq \ell_2$. Note that for $\ell_2$ to replicate an entry at index $i$, it must have a value $v'$ committed at entry $i-1$. Consider the last leader, $\ell_3$, who wrote a value on $\ell_2$'s $i-1$th entry. If $\ell_3 = \ell_1$, then $v' = v$, since a single leader only ever writes one value on each index. Thus, if agreement is broken by $p$ at time $t$, then it must have also been broken at an earlier time by $\ell_2$, which had $v$ committed at $i-1$ before time $t$. Contradiction.

If $\ell_3 = \ell_2$,
we consider two cases, depending on whether or not $p$ is part of $\ell_2$'s confirmed followers set. 
If $p$ is not in the confirmed followers of $\ell_2$, then $\ell_2$ could not have written a value on $p$'s $i$th log slot. Therefore, $p$ must have been a confirmed follower of $\ell_2$. If $p$ was part of $\ell_2$'s quorum for committing entry $i-1$, then $\ell_2$ was the last leader to write $p$'s $i-1$th slot, contradicting the assumption that $\ell_1$ wrote it last. Otherwise, if $\ell_2$ did not use $p$ as part of its quorum for committing, it still must have created a work request to write on $p$'s $i-1$th entry before creating the work request to write on $p$'s $i$th entry. By the FIFOness of RDMA queue pairs, $p$'s $i-1$th slot must therefore have been written by $\ell_2$ before the $i$th slot was written by $\ell_2$, leading again to a contradiction.

Finally, consider the case where $\ell_3 \neq \ell_1$ and $\ell_3 \neq \ell_2$. Recall from the previous case that $p$ must be in $\ell_2$'s confirmed followers set. Then when $\ell_2$ takes over as leader, it executes the update followers optimization presented in Algorithm~\ref{alg:app-updateFollowers}. By executing this, it must update $p$ with its own committed value at $i-1$, and update $p$'s FUO to $i$. However, this contradicts the assumption that $p$'s FUO was changed from $i-1$ to $i$ by $p$ itself using Algorithm~\ref{alg:FUO}.
\end{proof}

\subsubsection{Grow Confirmed Followers}

In our algorithm, the leader only writes to and reads from its confirmed followers set. So far, for a given leader $\ell$, this set is fixed and does not change after $\ell$ initially constructs it in lines~\ref{line:app-permission-start}--\ref{line:app-permission-end}. This implies that processes outside of $\ell$'s confirmed followers set will remain behind and miss updates, even if they are alive and timely.

We present an extension which allows such processes to join $\ell$'s confirmed followers set even if they are not part of the initial majority. Every time Propose is invoked, $\ell$ will check to see if it received permission acks since the last Propose call and if so, will add the corresponding processes to its confirmed followers set. This extension is compatible with those presented in the previous subsections: every time $\ell$'s confirmed followers set grows, $\ell$ re-updates its own log from the new followers that joined (in case any of their logs is ahead of $\ell$'s), as well as updates the new followers' logs (in case any of their logs is behind $\ell$'s).

One complication raised by this extension is that, if the number of confirmed followers is larger than a majority, then $\ell$ can no longer wait for its reads and writes to complete at all of its confirmed followers before continuing execution, since that would interfere with termination in an asynchronous system.

The solution is for the leader to issue reads and writes to all of its confirmed followers, but only wait for completion at a majority of them. One crucial observation about this solution is that confirmed followers cannot miss operations or have operations applied out-of-order, even if they are not consistently part of the majority that the leader waits for before continuing. This is due to RDMA's FIFO semantics.

The correctness of this extension derives from the correctness of the algorithm in general; whenever a leader $\ell$ adds some set $S$ to its confirmed followers $C$, forming $C' = C \cup S$, the behavior is the same as if $\ell$ just became leader and its initial confirmed followers set was $C'$.

\subsubsection{Omit Prepare Phase}

As is standard practice for Paxos-derived implementations, the prepare phase can be omitted if there is no contention. More specifically, the leader executes the prepare phase until it finds no accepted values during its prepare phase (i.e., until the check at line~\ref{line:app-checkEmpty} succeeds). Afterwards, the leader omits the prepare phase until it either (a) aborts, or (b) grows its confirmed followers set; after (a) or (b), the leader executes the prepare phase until the check at line~\ref{line:app-checkEmpty} succeeds again, and so on.

This optimization concerns performance on the common path. With this optimization, the cost of a Propose call becomes a single RDMA write to a majority in the common case when there is a single leader.

The correctness of this optimization follows from the following lemma, which states that no `holes' can form in the log of any replica. That is, if there is a value written in slot $i$ of process $p$'s log, then every slot $j<i$ in $p$'s log has a value written in it.

\begin{lemma}[No holes] \label{lem:noholes}
For any process $p$, if $p$'s log contains a value at index $i$, then $p$'s log contains a value at every index $j$, $0 \le j \le i$.
\end{lemma}

\begin{proof}
Assume by contradiction that the lemma does not hold. Let $p$ be a process whose slot $j$ is empty, but slot $j+1$ has a value, for some $j$. Let $\ell$ be the leader that wrote the value on slot $j+1$ of $p$'s log, and let $t$ be the last time at which $\ell$ gained write permission to $p$'s log before writing the value in slot $j+1$. Note that after time $t$ and as long as $\ell$ is still leader, $p$ is in $\ell$'s confirmed followers set. By Algorithm~\ref{alg:app-updateFollowers}, $\ell$ must have copied a value into all slots of $p$ that were after $p$'s FUO and before $\ell$'s FUO. By the way FUO is updated, $p$'s FUO cannot be past slot $j$ at this time. Therefore, if $\ell$'s FUO is past $j$, slot $j$ would have been populated by $\ell$ at this point in time. Otherwise, $\ell$ starts replicating values to all its confirmed followers, starting at its FUO, which we know is less than or equal to $j$. By the FIFO order of RDMA queue pairs, $p$ cannot have missed updates written by $\ell$. Therefore, since $p$'s $j+1$th slot gets updated by $\ell$, so must its $j$th slot. Contradiction.
\end{proof}

\begin{corollary}
Once a leader reads no accepted values from a majority of the followers at slot $i$, it may safely skip the prepare phase for slots $j>i$ as long as its confirmed followers set does not decrease to less than a majority.
\end{corollary}

\begin{proof}
Let $\ell$ be a leader and $C$ be its confirmed follower set which is a quorum. Assume that $\ell$ executes line~\ref{line:app-adoptOwn} for slot $i$; that is, no follower $p\in C$ had any value in slot $i$. Then, by Lemma~\ref{lem:noholes}, no follower in $C$ has any value for any slot $j>i$. Since this constitutes a majority of the processes, no value is decided in any slot $j>i$, and by Invariant~\ref{inv:commit}, no value is committed at any process at any slot $j>i$. Furthermore, as long as $\ell$ has the write permission at a majority of the processes, $\ell$ is the only one that can commit new entries in these slots (by Invariant~\ref{inv:solo-detection}). Thus, $\ell$ cannot break agreement by skipping the prepare phase on the processes in its confirm followers set.
\end{proof}

\fi
	
\end{document}